\documentclass[10pt]{article}
\usepackage{amssymb,latexsym, amsmath}
\newcommand{\bfi}{\bfseries\itshape}
\makeatletter
\@addtoreset{figure}{section}
\def\thefigure{\thesection.\@arabic\c@figure}
\def\fps@figure{h, t}
\@addtoreset{table}{bsection}
\def\thetable{\thesection.\@arabic\c@table}
\def\fps@table{h, t}
\@addtoreset{equation}{section}

\makeatother

\newtheorem{thm}{Theorem}[section]

\begin{document}
\pagestyle{myheadings}
\markright{ LAUR-97-3326\hspace{.5in}
{\it Maxwell-Vlasov system in Euler-Poincar\'{e} form}}

\title{The Maxwell-Vlasov equations\\ in Euler-Poincar\'{e} form}
\author{Hern\'an Cendra
\\Control and Dynamical Systems\\
California Institute of Technology 107-81\\ Pasadena, CA 91125
\\{\footnotesize uscendra@cco.caltech.edu}
\thanks{Permanent address: Universidad Nacional del Sur,
8000 Bahia Blanca, Argentina.
email: uscendra@criba.edu.ar}
\and
Darryl D. Holm
\thanks{Research supported by US DOE and the University of California}
\\Los Alamos National Laboratory\\ Los Alamos, NM 87545
\\ {\footnotesize dholm@lanl.gov}
\and
Mark J. W. Hoyle
\\Department of Mathematics
\\University of California, Santa Cruz, CA 95064
\\ {\footnotesize mjwh1@cats.ucsc.edu}
\and
Jerrold E. Marsden\thanks{Research partially supported by
DOE contract DE--FG0395--ER25251}
\\Control and Dynamical Systems\\
California Institute of Technology 107-81\\ Pasadena, CA 91125
\\ {\footnotesize marsden@cds.caltech.edu}\\
\\To appear in {\it J. Math. Phys.}\\}
\date{PACS Numbers: 03.40.Gc, 47.10.+g, 03.40.-t, 03.40.-z}
\maketitle

\begin{abstract}
Low's well known action principle for the Maxwell-Vlasov
equations of ideal plasma dynamics was originally expressed in terms
of a mixture of Eulerian and Lagrangian variables. By imposing
suitable constraints on the variations and analyzing invariance
properties of the Lagrangian, as one does for the Euler equations for
the rigid body and ideal fluids, we first transform this action
principle into purely Eulerian variables.  Hamilton's principle for
the Eulerian description of Low's action principle then casts the
Maxwell-Vlasov equations into Euler-Poincar\'{e} form for right
invariant motion on the diffeomorphism group of position-velocity
phase space,
$\mathbb{R}^{6}$. Legendre transforming the Eulerian form of Low's
action principle produces the Hamiltonian formulation of these
equations in the Eulerian description. Since it arises from
Euler-Poincar\'{e} equations, this Hamiltonian formulation can be
written in terms of a Poisson structure that contains the Lie-Poisson
bracket on the dual of a semidirect product Lie algebra. Because of
degeneracies in the Lagrangian, the Legendre transform is dealt with
using the Dirac theory of constraints.  Another Maxwell-Vlasov Poisson
structure is known, whose ingredients are the Lie-Poisson bracket on
the dual of the Lie algebra of symplectomorphisms of phase space and
the Born-Infeld brackets for the Maxwell field. We discuss the
relationship between these two Hamiltonian formulations. We also
discuss the general Kelvin-Noether theorem for Euler--Poincar\'e
equations and its meaning in the plasma context.
\end{abstract}
\tableofcontents

\section{Introduction} \label{sec-intro}

\paragraph{Reduction of action principles.} Due to their wide
applicability, the Maxwell-Vlasov equations of ideal plasma
dynamics have been studied extensively. Low [1958] wrote down an
action principle for them in preparation for studying stability of
plasma equilibria. Low's action principle is expressed in terms of
a mixture of Lagrangian particle variables and Eulerian field
variables.

Following the initiative of Arnold [1966] and its later
developments (see Marsden and Ratiu [1994] for background), we
start with a purely Lagrangian description of the plasma and
investigate the invariance properties of the corresponding action.
Using this set up and recent developments in the theory of the
Euler-Poincar\'e equations (Poincar\'{e} [1901b]) due to Holm, Marsden
and Ratiu [1997], we are able to cast Low's action principle into a
purely Eulerian description.

In this paper, we start with the {\it standard} form of Hamilton's
variational principle (in the Lagrangian representation) and {\it
derive} the new Eulerian action principle by a systematic reduction
process, much as one does in the corresponding derivation of Poisson
brackets in the Hamiltonian formulation of the Maxwell-Vlasov
equations starting with the {\it standard canonical brackets} and
proceeding by symmetry reduction (as in Marsden and Weinstein [1982]).
In particular, the Eulerian action principle we obtain in this way is
different from the ones found in Ye and Morrison [1992] by ad hoc
procedures. We also mention that the method of reduction of
variational principles we develop naturally justifies constraints on
the variations of the so called ``Lin constraint'' form, well known in
fluid mechanics.

The methods of this paper are based on reduction of variational
principles, that is, on Lagrangian reduction (see Cendra et al.
[1986, 1987] and Marsden and Scheurle [1993a,b]). These methods
have also been useful for systems with nonholonomic
constraints. This has been demonstrated in the work of  Bloch,
Krishnaprasad, Marsden and Murray [1996], who derived the reduced
La\-gran\-ge d'Al\-em\-bert equations for nonholonomic systems, which
also have a constrained variational structure. The methods of
the present paper should enhance the applicability of the Lagrangian
reduction techniques for even wider classes of continuum systems.

\paragraph{Passage to the Hamiltonian formulation.}
The Hamiltonian structure and nonlinear stability properties of the
equilibrium solutions for the Maxwell-Vlasov system have been
thoroughly explored. Some of the key references are Iw\'{i}nski and
Turski [1976], Morrison [1980], Marsden and Weinstein [1982] and Holm,
Marsden, Ratiu and Weinstein [1985]. See also the introduction and
bibliography of Marsden, Weinstein et al. [1983] for a guide to the
history and literature of this subject.

In our approach, Lagrangian reduction leads to the
Euler-Poincar\'{e} form of the equations, which is still in the
Lagrangian formulation. Using this set up, one may pass from the
Lagrangian to the Hamiltonian formulation of the Maxwell-Vlasov
equations by Legendre transforming the action principle in the Eulerian
description at either the level of the group variables (the level
that keeps track of the particle positions), or at the level of
the Lie algebra variables. One must be cautious in this procedure
because the relevant Hamiltonian and Lagrangian are degenerate.
We deal with this degeneracy by using a version of the Dirac theory of
constraints.

Legendre transforming at the group level leads to a canonical
Hamiltonian formulation and the latter leads to a new Hamiltonian
formulation of the Maxwell-Vlasov equations in terms of a Poisson
structure containing the Lie-Poisson bracket on the dual of a
semidirect product Lie algebra. This new formulation leads us
naturally to the starting point for Hamiltonian reduction used by
Marsden and Weinstein [1982] (see also Morrison [1980] and Kaufman and
Dewar [1984]).

\paragraph{Stability and asymptotics.} The new Hamiltonian formulation
of the Maxwell-Vlasov system places these equations into a
framework in which one can use the energy-momentum and
energy-Casimir methods for studying nonlinear stability properties of
their relative equilibrium solutions. This is directly in line with
Low's intended program, since the study of stability was Low's original
motivation for writing his action principle. Sample references in
this direction are Holm, Marsden, Weinstein and Ratiu [1985],
Morrison [1987], Morrison and Pfirsch [1990], Wan [1990], Batt and
Rein [1993] and Batt, Morrison and Rein [1995].  Other historical
references for the Lagrangian approach to the Maxwell-Vlasov equations
include Sturrock [1958], Galloway and Kim [1971] and Dewar [1972].

The Eulerian formulation of
Low's action principle also casts it into a form that is amenable to
asymptotic expansions and creation of approximate theories (such as
guiding center theories) possessing the same mathematical structure
arising from the Euler-Poincar\'{e} setting. See, for example, Holm
[1996] for applications of this approach of Hamilton's principle
asymptotics in geophysical fluid dynamics.

\paragraph{Comments on the Maxwell-Vlasov system.} The rest of this
paper will be concerned with variational principles for the
Maxwell-Vlasov system of equations for the dynamics of an ideal
plasma. These equations have a long history dating back at least to
Jeans [1902], who used them in a simpler form known as the
Poisson-Vlasov system to study structure formation on stellar and
galactic scales. Even before Jeans, Poincar\'e [1890, 1901a] had
investigated the stability of equilibrium solutions of the
Poisson-Vlasov system for the purpose of determining the stability
conditions for steller configurations. The history of the efforts to
establish stellar stability conditions using the Poisson-Vlasov system
is summarized by Chandrasekhar [1977]. The Poisson-Vlasov system is
also used to describe the self-consistent dynamics of an electrostatic
collisionless plasma, whereas the Maxwell-Vlasov system is used to
describe the dynamics of a collisionless plasma evolving
self-consistently in an electromagnetic field.

\paragraph{Organization of the paper.}
The paper is organized as follows. Section \ref{sec-Maxwell}
introduces the Maxwell-Vlasov equations. In section \ref{sec-EP} we
state the Euler-Poincar\'{e} theorem for Lagrangians depending on
parameters along with the associated Kelvin-Noether theorem. This
general theorem plays a key role in our analysis. Section
\ref{sec-MVEP} reformulates these equations in a purely Eulerian form
and shows how they satisfy the Euler-Poincar\'{e} theorem. The
following section reviews some aspects of the Legendre transformation
for degenerate Lagrangians. Section \ref{sec-action} reprises Low's
action principle for the Maxwell-Vlasov equations. Section
\ref{sec-Ham} casts the Euler-Poincar\'{e} formulation of the
Maxwell-Vlasov equations into Hamiltonian form possessing a Poisson
structure that contains a Lie-Poisson bracket. In Section
\ref{sec-conc} we summarize our conclusions.

\section{The Maxwell-Vlasov equations} \label{sec-Maxwell}

The Maxwell-Vlasov system of equations describes the single
particle distribution for a set of charged particles of one
species moving self-consistently in an electromagnetic field. In this
description, the Boltzmann function $f({\bf x},{\bf v},t)$ is
viewed as the instantaneous probability density function for the
particle distribution, i.e., given a region $\Omega $ of phase
space, the probability of finding a particle in that region is

\begin{equation}\label{prob.vlasov}
\int_{\Omega}d{\bf x}\, d{\bf v}f({\bf x},{\bf v},t),
\end{equation}
where {\bf x} and {\bf v} are the current positions and velocities of
the plasma particles. Thus, if the phase-space domain $\Omega$ is the
whole ({\bf x},{\bf v}) space, the value of this integral at a certain
time $t$ is normalized to unity.

As is customary, we assume that the particles of the plasma
obey dynamical equations and that the plasma density $f$ is
advected as a scalar along the particle trajectories in phase space,
i.e.,
\begin{equation}
\frac{\partial f}{\partial t}
+{\dot{\bf x}}\cdot \nabla _{\bf x}f
+{\dot{\bf v}}\cdot \nabla _{\bf v}f=0.
\label{eq:pv1}
\end{equation}
In this equation, and in the sequel, an
overdot refers to a time derivative along a phase space trajectory,
and $\nabla_{\bf x}$ and $\nabla _{\bf v}$ denote the gradient
operators with respect to position and velocity respectively.
For pressureless motion in the
electromagnetic field of the charged particle distribution, the
acceleration of a particle is given by
\begin{equation}\label{acc.eqn}\ddot{\bf x}
=-\,\frac{q }{m } \left[\nabla_{\bf x} \Phi+
\frac{\partial {\bf A}}{\partial t}
-{\bf v}\times(\nabla_{\bf x}\times {\bf A})\right],
\end{equation}
where $(q/m)$ denotes the charge to mass ratio of an individual
particle, $\Phi$ is the electric potential, and ${\bf A}$ is the
magnetic vector potential. Substituting this expression for
$\dot{\bf v}$ in equation (\ref{eq:pv1}) yields
\begin{equation}
\frac{\partial f}{\partial t}
+{\bf v}\cdot \nabla_{\bf x} f
-\frac{q}{m}\bigg[\nabla _{\bf x} \Phi
+\frac{\partial {\bf A}}{\partial t}
-{\bf v}\times(\nabla_{\bf x}\times {\bf A})\bigg]
\cdot \nabla _{\bf v}f=0\,.
\label{eq:CBE}
\end{equation}
This is the {\bfi Vlasov equation} (also called
the collisionless Boltzmann, or Jeans equation). The system is
completed by the Maxwell equations with sources:
\begin{equation}
\nabla_{\bf x}\cdot {\bf E}=\rho,
\quad \quad
\nabla_{\bf x} \times {\bf B}
=\frac{\partial {\bf E}}{\partial t}+{\bf j},
\label{Maxwell}
\end{equation}
where {\bf E} and {\bf B} are the electric and magnetic field
variables respectively, $\rho$ is the charge density and {\bf j}
is the current density. These quantities are expressed
in terms of the Boltzmann function $f$ and the Maxwell scalar and
vector potentials $\Phi$ and {\bf A} by:
\begin{eqnarray}
{\bf E}=-\nabla_{\bf x} \Phi-\frac{\partial {\bf A}}{\partial t},
& {\bf B}=\nabla_{\bf x}\times{\bf A},
\nonumber \\
\rho({\bf x},t)=q \int d{\bf v}\, f({\bf x},{\bf v},t),
&
{\displaystyle {\bf j}({\bf x},t)
=q\int d{\bf v}\,
{\bf v}f({\bf x},{\bf v},t).}
\label{rhoandj}
\end{eqnarray}
By their definitions, {\bf E} and {\bf B} satisfy the kinematic
Maxwell equations
\begin{equation}
\nabla_{\bf x}\cdot{\bf B}=0,
\qquad\nabla_{\bf x} \times {\bf E}=
-\,\frac{\partial {\bf B}}{\partial t}.
\label{Maxwell2}
\end{equation}

Equations (\ref{eq:CBE}) - (\ref{Maxwell2}) comprise the {\bfi
Maxwell-Vlasov equations}. When {\bf A} is absent, the field is
electrostatic and one obtains the Poisson-Vlasov equations. The
Poisson-Vlasov system can also be used to describe a self gravitating
collisionless fluid, and so it forms a model for the evolution of
galactic dynamics, see, e.g., Binney and Tremaine [1987].

Note that the integral in (\ref{prob.vlasov}) is independent of time
(as the region and the function $f$ evolve), since the
vector field defining the motion of particles (see equation
(\ref{acc.eqn})) is divergence free with respect to the standard
volume element on velocity phase space. Thus, one may interpret $f$
either as a density or as a scalar. For our purposes later,
we will need to be careful with the distinction, since the volume
preserving nature of the flow of particles will be a consequence of
our variational principle and will not be imposed at the outset.

\section{The Euler-Poincar\'e equations, Semidirect Products, and
Kelvin's Theorem}\label{sec-EP}
\paragraph{The general Euler-Poincar\'e equations.}
Here we recall from Holm, Marsden and Ratiu [1997] the general form of
the Euler-Poincar\'e equations and their associated Kelvin-Noether
theorem. In the next section, we will immediately specialize these
statements for a general invariance group $G$ to the case of plasmas
when $G$ is the diffeomorphism group, ${\rm Diff}(T\mathbb{R}^3)$. We
shall state the general theorem for right actions and right invariant
Lagrangians, which is appropriate for the Maxwell-Vlasov situation.
The notation is as follows.

\begin{itemize}
\item There is a {\it right\/} representation of the Lie group $G$ on
the vector space $V$ and $G$ acts in the natural way from the {\it
right\/} on $TG \times V^\ast$: $(v_g, a)h = (v_gh, ah)$.

\item $\rho _v: \mathfrak{g}  \rightarrow V$ is the linear map
given by the corresponding right action of the Lie algebra on $V$:
$\rho_v(\xi) = v \xi$, and $\rho _v^\ast: V ^\ast  \rightarrow
\mathfrak{g}^\ast$ is its dual. The $\mathfrak{g} $--action on
$\mathfrak{g} ^\ast$ and $V^\ast$ is defined to be {\it minus} the
dual map of the $\mathfrak{g} $--action on $\mathfrak{g} $ and $V$
respectively and is denoted by $\mu \xi$ and $a \xi$ for $\xi \in
\mathfrak{g} $, $\mu \in \mathfrak{g} ^\ast$, and $a\in V^\ast$. For
$v \in V $ and $a \in V ^\ast$, it will be convenient to write:
\[
v \diamond  a = \rho _v ^\ast a  \quad {\rm i.e., }
\quad \left\langle v \diamond  a, \xi \right\rangle
=   \left\langle a, v \xi \right\rangle
= - \left\langle v, a \xi \right\rangle\,,
\]
for all $\xi \in \mathfrak{g}$. Note that $v \diamond  a \in
\mathfrak{g}^\ast$.
\item Let ${\mathcal Q}$ be a manifold on which $G$ acts {\it
trivally} and assume that we have a function $L : T G \times
T{\mathcal Q}  \times V ^\ast \rightarrow \mathbb{R}$ which is
right $G$--invariant.
\item In particular, if $a_0 \in V^\ast$, define the Lagrangian
$L_{a_0} : TG \times T{\mathcal Q} \rightarrow \mathbb{R}$ by
$L_{a_0}(v_g, u _q) = L(v_g, u _q, a_0)$. Then $L_{a_0}$ is right
invariant under the lift to $TG \times T{\mathcal Q}$ of the right
action of $G_{a_0}$ on $G \times {\mathcal Q}$.
\item  Right $G$--invariance of $L$ permits us to define
$l: \mathfrak{g} \times T{\mathcal Q}
\times V^\ast \rightarrow \mathbb{R}$ by
\[
l(v_gg^{-1}, u _q, ag^{-1}) = L(v_g, u _q,  a).
\]
Conversely,  this relation defines for any
$l: \mathfrak{g} \times T{\mathcal Q} \times V^\ast \rightarrow
\mathbb{R}$ a right $G$--invariant function
$ L : T G  \times T{\mathcal Q} \times V ^\ast
\rightarrow \mathbb{R} $.
\item For a curve $g(t) \in G, $ let
$\xi(t) := \dot{g}(t) g(t)^{-1}$ and define the curve
$a(t)$ as the unique solution of the linear differential equation
with time dependent coefficients $\dot a(t) = -a(t)\xi(t)$
with initial condition $a(0) = a_0$. The solution can be
equivalently written as $a(t) = a_0g(t)^{-1}$.
\end{itemize}

\begin{thm} \label{rarl}
The following are equivalent:
\begin{enumerate}
\item [{\bf i} ] Hamilton's variational principle holds:
\begin{equation} \label{hamiltonprincipleright1}
\delta \int _{t_1} ^{t_2}
L_{a_0}(g(t), \dot{g} (t), q (t), \dot{q} (t) )
dt = 0\,,
\end{equation}
for variations of $g$ and $q $ with fixed endpoints.
\item [{\bf ii}  ] $\left(g(t), q (t)\right)$ satisfies
the Euler--Lagrange equations for $L_{a_0}$ on $G \times {\mathcal
Q}$.
\item [{\bf iii} ]  The constrained variational
principle\footnote{Strictly speaking this is not a variational
principle because of the constraints imposed on the variations.
Rather, this principle is more like the La\-gran\-ge d'Al\-em\-bert
principle used in nonholonomic mechanics.}
\begin{equation} \label{variationalprincipleright1}
\delta \int _{t_1} ^{t_2}
l(\xi(t), q (t), \dot{q}(t),  a(t)) dt = 0\,,
\end{equation}
holds on $\mathfrak{g} \times {\mathcal Q}$, upon using variations of
the form
\begin{equation} \label{variationsright1}
\delta \xi
= \frac{\partial \eta }{\partial t} - ad_{\xi} \eta
= \frac{\partial \eta }{\partial t} - [\xi , \eta ],
\quad
\delta a =  -a\eta ,
\end{equation}
where $\eta(t) \in \mathfrak{g}$ vanishes at the endpoints and
$\delta q (t)$ is unrestricted except for vanishing at the endpoints.
\item [{\bf iv}] The following  system of Euler--Poincar\'{e}
equations (with a parameter) coupled with Euler-Lagrange equations
holds on $\mathfrak{g} \times T{\mathcal Q} \times V^\ast$:
\begin{equation} \label{eulerpoincareright1}
 \frac{\partial }{\partial t}
 \frac{\delta l}{\delta \xi} = -
\mbox {\rm ad}_{\xi}^{\ast} \frac{ \delta l }{ \delta \xi}
+ \frac{\delta l }{ \delta a} \diamond  a\,,
\end{equation}
and
\begin{equation}
 \frac{\partial }{\partial t}
 \frac{\partial l }{\partial \dot{q}^i}
- \frac{\partial l}{\partial q ^i} = 0\,.
\end{equation}
\end{enumerate}
\end{thm}

The strategy of the proof is simple: one just determines the form of
the variations on the reduced space $\mathfrak{g} \times {\mathcal Q}
\times V ^\ast$ that are induced by variations on the unreduced
space $TG\times T{\mathcal Q}$ and includes the relation of
$a(t)$ to $a_0$. One then carries the variational principle to the
quotient. See Holm, Marsden and Ratiu [1997] for details. Here we have
included the extra factor of ${\mathcal Q}$ which is needed in the
present application; this will be the space of potentials for the
Maxwell field. This extra factor does not substantively alter the
arguments.

\paragraph{The Kelvin-Noether Theorem.}  We start with a Lagrangian
$L _{a _0}$ depending on a parameter $a _0 \in V ^\ast$ as above and
introduce a manifold ${\mathcal C}$ on which $G$ acts. We assume this
is also a right action and suppose we have an equivariant map
$\mathcal{K} : {\mathcal C}  \times V ^\ast
\rightarrow \mathfrak{g} ^{\ast \ast} $.

In the case of continuum theories, the space ${\mathcal
C}$ is chosen to be a loop space and $\left\langle \mathcal{K} (c,
a), \mu\right\rangle$ for $c \in {\mathcal C}$ and $\mu \in
\mathfrak{g}^\ast$ will be  a circulation. This class of examples
also shows why we {\it do not} want to identify the double dual
$\mathfrak{g} ^{\ast \ast}$ with $\mathfrak{g}$.

Define the {\bfi Kelvin-Noether quantity}
$I : {\mathcal C}  \times \mathfrak{g} \times T{\cal Q} \times V ^\ast
\rightarrow \mathbb{R}$ by
\begin{equation}\label{KelvinNoether}
I(c, \xi, q, \dot{q}, a) = \left\langle\mathcal{K} (c, a),
\frac{\delta l}{\delta \xi}( \xi , q, \dot{q} , a)
\right\rangle.
\end{equation}

\begin{thm}[Kelvin-Noether] \label{KelvinNoetherthm}Fixing $c_0 \in
{\mathcal C}$, let $\xi (t), q(t) , \dot{q}(t) , a(t)$ satisfy the
Euler-Poincar\'e equations and define $g(t)$ to be the solution of
$\dot{g}(t) = \xi(t) g(t)$ and, say, $g(0) = e$. Let
$c(t) = g(t)^{-1} c_0$ and $I(t)
= I(c(t), \xi(t), q(t) , \dot{q}(t) ,a(t))$.
Then
\begin{equation}
\frac{d}{dt} I(t) = \left\langle \mathcal{K}(c(t), a (t)
),\frac{\delta l}{\delta a} \diamond a
\right\rangle.
\label{Kel-Noeth}
\end{equation}
\end{thm}

The proof of this theorem is relatively straightforward; we refer to
Holm, Marsden and Ratiu [1997]. We shall express the relation
(\ref{Kel-Noeth}) explicitly for Maxwell-Vlasov plasmas at the end of
section \ref{sec-Ham}.

\section{An action for the Maxwell-Vlasov equations}
\label{sec-action}
A typical element of $T\mathbb{R} ^3 \cong \mathbb{R} ^3 \times
\mathbb{R} ^3$ will be denoted ${\bf z} = ({\bf x}, {\bf v})$. We let
$\pi_{s} : T\mathbb{R}^3 \rightarrow \mathbb{R}^3$
and $\pi_{v} : T\mathbb{R}^3 \rightarrow \mathbb{R}^3$
be  the projections $\pi_{s}({\bf z}) = {\bf x}$ and $\pi_{v}({\bf z})
= {\bf v}$ onto the first and second factors, respectively.

\paragraph{Spaces of fields.} We  let ${\rm Diff}(T\mathbb{R}^3)$
denote the group of
$C^{\infty}$- diffeomophisms from $T\mathbb{R}^3$ onto itself. An
element $\psi \in {\rm Diff}(T\mathbb{R}^3)$ maps plasma particles
having initial position and velocity $({\bf x}_0, {\bf v}_0)$ to their
current position and velocity
$({\bf x}, {\bf v}) = \psi({\bf x}_0, {\bf v}_0)$. This is the
particle evolution map. We shall sometimes abbreviate $({\bf x}_0,
{\bf v}_0) = {\bf z}_0$, $({\bf x},{\bf v}) = {\bf z}$, etc. The
spatial components of
$\psi({\bf x}_0, {\bf v}_0)$ are written as ${\bf x}( {\bf x}_0, {\bf
v}_0) $ and the velocity components as ${\bf v}( {\bf x}_0, {\bf
v}_0)$. We shall also use the following notation:
\begin{itemize}
\item
${\cal V} = C^{\infty}(\mathbb{R}^3, \mathbb{R})$ is the space of
electric potentials $\Phi({\bf x})$;
\item ${\cal A}$ is the space of magnetic potentials $A({\bf x})$;
\item ${\cal F} = C^{\infty}(T\mathbb{R}^3, \mathbb{R})$ is the
space of plasma densities  $f({\bf x},{\bf v})$;
\item ${\cal F}_0 = C^{\infty}_0(T\mathbb{R}^3, \mathbb{R})$ is the
space of plasma densities with  compact support;
\item ${\cal D}_0 = C^{\infty}_0(\mathbb{R}^3, \mathbb{R})$ is a
space of test functions, denoted $\varphi( {\bf x})$.
\end{itemize}

The test functions $\varphi({\bf x})$ are used to localize the
variational principle. Thus, once one obtains Euler-Lagrange equations
depending on $f_0$ and $\varphi_0$, if their validity can
be naturally extended for any $f_0$ and $\varphi_0$, which will
happen in our case, then we shall consider those extended equations
to be the Euler-Lagrange equations of the system. We will usually be
interested in the Euler-Lagrange equations for $f_0 >0$ and
$\varphi_0 = 1$.

\paragraph{The Lagrangian and the action.} For each choice of the
initial plasma distribution function $f_0$ and the test function
$\varphi_0$, we define the Lagrangian
\begin{eqnarray}
 L_{f_0,\varphi_0}(\psi,\dot{\psi},
\Phi, \dot{\Phi}, {\bf A}, \dot{{\bf A}})
& = &
\int d{\bf x}_{0}d{\bf v}_{0} f_0({\bf x}_{0},{\bf v}_{0})
 \left(\frac{1}{2}m|\dot{\bf x}({\bf x}_{0},{\bf v}_{0})|^{2}\right.
   \nonumber \\
& &
+ \,  \frac{1}{2}m|\dot{\bf x}({\bf x}_{0},{\bf v}_{0})
- {\bf v}({\bf x}_{0},{\bf v}_{0})|^{2}
\label{eq:Low} \\
& &
+ \left. \phantom{\frac{1}{2}}q\dot{\bf x}({\bf x}_{0},{\bf v}_{0})
\cdot {\bf A}({\bf x}({\bf x}_0, {\bf v}_0))
- q\Phi({\bf x}({\bf x}_0, {\bf v}_0))
\right)
\nonumber \\
&&
+\; \frac{1}{2}\int d{\bf r} \, \varphi _0 ( {\bf r})
\left(|\nabla_{\bf r}\Phi
+ \frac{\partial{\bf A}}{\partial t}({\bf r})|^{2}
-|\nabla_{\bf r} \times {\bf A}({\bf r})|^{2}\right).
\nonumber
\end{eqnarray}
This Lagrangian is the natural generalization of that for an
$N$-particle system, with terms corresponding to kinetic energy,
electric and magnetic field energy, the usual magnetic coupling term
with coupling constant $q$ (the electric charge), and a constraint
that ties the Eulerian fluid velocity ${\bf v}$ to
$\dot{\bf x}$, the material derivative of the Lagrangian particle
trajectory. Here ${\bf x}$ and
${\bf v}$ are Lagrangian phase space variables, while ${\bf A}$ and
$\Phi$ are Eulerian field variables. Thus, there should be no confusion
created by the slight abuse of notation in abbreviating ${\partial{\bf
A}}/{\partial t}$ and
${\partial\Phi}/{\partial t}$ as $\dot{\Phi}$ and $\dot{{\bf A}}$,
respectively, in the arguments of the Lagrangian. This Lagrangian is
inspired by Low [1958]. However, we have added the term
\[
\frac{1}{2}m|\dot{\bf x}({\bf x}_{0},{\bf v}_{0})
-{\bf v}({\bf x}_{0},{\bf v}_{0})|^{2}\,,
\]
which allows ${\bf v}$ to be varied independently in the variational
treatment.

Consider the action
\[
\mathfrak{S}
= \int dt\  L_{f_0,\varphi_0}(\psi,\dot{\psi},
\Phi, \dot{\Phi}, {\bf A}, \dot{{\bf A}})\,,
\]
defined on the family of curves $\left(\psi(t), \Phi(t),
{\bf A}(t)\right)$   satisfying the usual fixed-endpoint conditions
$\left(
\psi(t_{i}), \Phi(t_{i}), {\bf A}(t_{i})
\right) =
\left(
\psi_{i}, \Phi_{i}, {\bf A}_{i}
\right)$,
$i = 1, 2$. One now applies the standard techniques of the calculus of
variations. In particular, integration by parts can be performed
since $f_0$ and $\varphi_0$ have compact support. Moreover, once the
Euler-Lagrange equations have been obtained, their validity can be
easily extended in a natural way for $f_0 > 0$ and $\varphi_0 = 1$.

\paragraph{Derivation of the equations.} To write the equations of
motion, we need some additional notation. Consider the evolution map
$\psi _{t}({\bf x}_{0},{\bf v}_{0})=({\bf x},{\bf v})$
so that $\psi _{t}$ relates the initial positions and velocities
of fluid particles to their positions and velocities at time $t$.
Let ${\bf u}$ be the corresponding vector field:
$${\bf u}({\bf x},{\bf v}) :=
\dot{\psi} _{t} \circ \psi _t ^{-1} ({\bf x},{\bf v}) =:
\dot{\bf x}\frac{\partial}{\partial {\bf x}}+
\dot{\bf v}\frac{\partial}{\partial {\bf v}}\,,$$
so the components of ${\bf u}$ are $(\dot{\bf x},\dot{\bf v})$.
Recall that the transport of $f _0$ as a scalar is given by
$f ( {\bf x}, {\bf v}, t )
= f _0 \circ \psi _t ^{-1}( {\bf x}, {\bf v})$,
which satisfies
\begin{equation}
\frac{\partial f}{\partial t}
+{\bf u}
\cdot \nabla_{\bf z}f=0,
\label{eff}
\end{equation}
where $\nabla_{\bf z}=(\nabla _{\bf x},\nabla _{\bf v})$ is the
six dimensional gradient operator in $({\bf x},{\bf v})$ space.
Let $J _\psi$ be the Jacobian determinant of the mapping $\psi \in
{\rm Diff}(T\mathbb{R}^3)$, that is, the determinant of the Jacobian
matrix $\partial({\bf x},{\bf v})/\partial ({\bf x}_0,{\bf v}_0)$.

Define $F({\bf x},{\bf v}, t)$ to be $f _0$, transported as a {\it
density}:
\[
F({\bf x}({\bf x}_0,{\bf v}_0),{\bf v}({\bf x}_0,{\bf v}_0), t)
J _\psi ({\bf x}_0,{\bf v}_0) =  f_0({\bf x}_0,{\bf v}_0),
\]
so that
\begin{equation}
\frac{\partial F}{\partial t}
+\nabla_{\bf z}
\cdot (F{\bf u})=0.
\label{useful}
\end{equation}
Taking variations in our Lagrangian (\ref{eq:Low}) and making use
of the preceding equation for $F$, we obtain the following equations
(taking $\varphi_0=1$)
\begin{equation}\label{eq:Pois}
\begin{array}{rc} \delta {\bf x}: &
{\displaystyle m\ddot{\bf x}+m(\ddot{\bf x}-\dot{\bf v})
=-q\nabla _{\bf x} \Phi -
q\frac{\partial {\bf A}}{\partial t}
+q\dot{\bf x}\times(\nabla_{\bf x} \times {\bf A})}, \\[6pt]
\delta {\bf v}: &
\dot{\bf x} -{\bf v}=0,\\[6pt]
\delta \Phi: & {\displaystyle \nabla _{\bf x}\cdot \left(
\nabla_{\bf x} \Phi +\frac{\partial {\bf A}}{\partial t}\right)
=-q\int \,d{\bf v}
F({\bf x},{\bf v},t)}, \\[9pt]
\delta {\bf A}: &
{\displaystyle \nabla_{\bf x} \times (\nabla_{\bf x} \times {\bf A})=
-\frac{\partial }{\partial t} \left(\nabla_{\bf x} \Phi
+\frac{\partial {\bf A}}{\partial t}\right)+q\int
d{\bf v}\,{\bf v}F({\bf x},{\bf v},t)}.
\end{array}
\end{equation}
The second equation in (\ref{eq:Pois}) treats the Eulerian
velocity ${\bf v}$ as a Lagrange multiplier, and ties its value to
the fluid velocity $\dot{\bf x}$, hence $\dot{\bf v}=\ddot{\bf x}$
as well. The first two variational equations in the set
(\ref{eq:Pois}) provide the desired relation for
particle acceleration and the last two equations are the
Maxwell equations with source terms. Thus, Hamilton's
principle with Low's action provides the equations for
self-consistent particle motion in an electromagnetic field, as
required, and the description is completed by substituting
\[ \left( {\bf v}, -\frac{q}{m}\left[\nabla _{\bf x} \Phi
+\frac{\partial{\bf A}}{\partial t}
-{\bf v}\times(\nabla_{\bf x}\times{\bf A})\right] \right) \]
for the components of ${\bf u}$
in the transport equation
(\ref{eff}) to give the Vlasov equation (\ref{eq:CBE}).

\section{The Maxwell-Vlasov system as Euler-Poincar\'{e} equations}
 \label{sec-MVEP}
We will now specialize the general Euler-Poincar\'e theorem to the
case of plasmas. The Lagrangian $L_{f_0,\varphi_0}(\psi,\dot{\psi},
\Phi, \dot{\Phi}, {\bf A},\dot{{\bf A}})$ in equation (\ref{eq:Low})
has a right ${\rm Diff}(T\mathbb{R}^3)$- symmetry. Let $\eta
\in {\rm Diff}(T\mathbb{R}^3)$, $F \in {\cal F}$ and define the
action of $\eta$ on $F$ by
$F\eta = (F \circ \eta)J_\eta$ where, as above, $J_\eta$ is
the Jacobian determinant of $\eta$.

The symmetry of
$L_{f_0,\varphi_0}(\psi,\dot{\psi}, \Phi, \dot{\Phi},{\bf A}, \dot{{\bf
A}})$ is the property
$$
L_{f_0\eta,\varphi_0}(\psi \eta,\dot{\psi} \eta, \Phi, \dot{\Phi},
{\bf A}, \dot{{\bf A}}) = L_{f_0,\varphi_0}(\psi,\dot{\psi}, \Phi,
\dot{\Phi}, {\bf A},\dot{{\bf A}}),
$$
for all $\eta \in {\rm Diff}(T\mathbb{R}^3)$.

\paragraph{Ingredients for Euler-Poincar\'e.} Now we apply the general
Euler-Poincar\'{e} theorem
\ref{rarl}, taking
$G = {\rm Diff}(T\mathbb{R}^3)$ and
${\mathcal Q} = {\mathcal V}\times{\mathcal A}$ and the parameter
$a_0 = f_0$. As we have explained before, $\varphi_0$ is an auxiliary
quantity that will ultimately take the value unity.
In the general Euler-Poincar\'{e} theorem \ref{rarl}
we take
\begin{equation}\label{tensor.eqn}
\delta {\bf u} =
  \frac{\partial {\bf w}}{\partial t}
  - {\rm ad}_{\bf u}{\bf w},\qquad
\delta a=-\pounds _{\bf w}a,
\end{equation}
where ${\bf w}\in \mathfrak{g}$ is a vector field on $T \mathbb{R}^3$,
$\pounds_{\bf w}$ is the Lie derivative and
${\rm ad}_{\bf u}{\bf w}=-[{\bf u},{\bf w}]$ defines
${\rm ad}_{\bf u}{\bf w}$
in terms of the Lie bracket of vector fields, $[{\bf u},{\bf w}]$.
The Euler-Poincar\'{e} equations (\ref{eulerpoincareright1}) are
\begin{equation}
\frac{\partial }{\partial t}\frac{\delta l}{\delta {\bf u}}=
 - ad^{\ast}_{\bf u}\frac{\delta l}{\delta {\bf u}}+
\frac{\delta l}{\delta a} \diamond  a,
\label{eq:EP}
\end{equation}
where ${\rm ad}^{\ast}_{\bf u}$ is the dual of  ${\rm ad}_{\bf
u}$ and ${\delta l}/{\delta {\bf u}}$ is a 1-form density. The 1-form
density $(\delta l/\delta a) \diamond  a$ is defined by
\begin{equation}
\left\langle \frac{\delta l}{\delta a}
\diamond  a, {\bf w} \right\rangle
= -\int \frac{\delta l}{\delta a}
\cdot \pounds_{\bf w} a.
\label{adj}
\end{equation}
When the quantities $a$ are tensor fields,
$\delta l/\delta a$ will be elements of the dual space under the
natural pairing.

We shall apply this result to obtain the Maxwell-Vlasov system
(\ref{eq:CBE})-(\ref{Maxwell2}) as Euler-Poincar\'e equations.
We begin by recording a formula that will be needed later. Let
${\bf u},{\bf w}$ be two elements of $\mathfrak{g}$, the Lie algebra
of vector fields for the diffeomorphism group on a manifold
$\mathcal{M}  $. Choose the 1-form density ${\bf c}\in
\mathfrak{g}^{\ast}$, and let the pairing
$\left\langle {\bf c},{\bf u}\right\rangle :\mathfrak{g}^{\ast}\times
\mathfrak{g}
\rightarrow \mathbb{R}$ be given by:
\begin{equation}
\left\langle {\bf c},{\bf u} \right\rangle =\int _\mathcal{M}  d{\bf z} \,{\bf
c}\cdot {\bf u} = \int _\mathcal{M}  d{\bf z} \,c_j u^j,
\label{eq:metric} \end{equation}
where $c_j$ and $u^j$, $j=1\dots n$, are components of ${\bf c}$ and
${\bf u}$ in $\mathbb{R}^n$ and $d{\bf z}$ is the volume form on
$\mathcal{M}$. Then we can write the desired formula,
\begin{eqnarray}
\left\langle {\rm ad}_{\bf  u}^{\ast}{\bf  c},{\bf w} \right\rangle
  & = &
\int d{\bf z} \,{\rm ad}_{\bf  u}^{\ast}{\bf  c}\cdot{\bf w}
  =
\int d{\bf z}\,{\bf c}\cdot {\rm ad}_{\bf u}{\bf w} \nonumber \\
  & = & - \int d{\bf z} \,c_{i}
\left( u^{j}\frac{\partial w^{i}}{\partial z^{j}}-
w^{j}\frac{\partial u^{i}}{\partial z^{j}} \right) \nonumber \\
  & = & \int d{\bf z}\, w^{i}
\left(c_{j}\frac{\partial u^{j}}{\partial z^{i}}
+ c_{i}( \nabla\cdot{\bf u}) + ({\bf u}\cdot\nabla )c_{i}\right)
\nonumber \\
  & = & \left\langle \pounds_{\bf  u}{\bf  c},{\bf w}\right\rangle .
\label{eq:ad1}
\end{eqnarray}
Here $\pounds_{\bf  u}{\bf  c}$ is the Lie derivative of the the
1-form density ${\bf c}$ with respect to the vector field ${\bf u}$,
$z^{j}$ is the coordinate chart and
$c_{j}, u^{j}, w^{j}$ are the components of vectors in
$\mathbb{R}^{n}$. Unless otherwise stated, we sum
repeated indices over their range, $i,j=1,\ldots,n$ where $n$ is the
dimension of
$\mathcal{M} $. We assume that the vector fields and 1-form densities
are defined so that integration by parts gives no contribution at the
boundary (inclusion of nonzero boundary terms is
straightforward). Formula (\ref{eq:ad1}) for ${\rm ad}_{\bf
u}^{\ast}{\bf c}$ will be useful later.

By definition,
${\bf u}=(\dot{\bf x},\dot{\bf v})$; we will denote
${\bf u}_{s}= \dot{\bf x}$, the spatial part
of the phase space velocity field.

\paragraph{The reduced action.} We may transform the action
(\ref{eq:Low}) into the Eulerian description as the reduced action
\begin{eqnarray}
\mathfrak{S}_{\rm red}   & = &  \int dt \, l ( {\bf u}, \Phi, \dot{
\Phi}, {\bf A} ,
\dot{{\bf A}} ) \nonumber \\  & = &
\int dt \, \int d{\bf x}d{\bf v}F({\bf x},{\bf v},t)
\left(\frac{1}{2}m|{\bf u}_{s}|^{2}
+\frac{1}{2}m|{\bf u}_{s}-{\bf v}|^{2}-
q\Phi +q{\bf u}_{s}\cdot{\bf A}\right) \nonumber \\
& & +  \ \frac{1}{2}\int dt\int d{\bf x}\ \Big|\nabla _{\bf x} \Phi
+\frac{\partial {\bf A}}{\partial t} \Big|^{2}
- |\nabla_{\bf x}\times {\bf A}|^{2}.
\label{eq:newlag}
\end{eqnarray}
We vary this action with respect to ${\bf u}_{s}$,
$F$, $\Phi$ and ${\bf A}$:
\begin{eqnarray}
\delta \mathfrak{S}_{\rm red} & = & \int dt\int d{\bf x}d{\bf
v}\bigg\{ F \bigg[\Big(m{\bf u}_{s}+m({\bf u}_{s}-{\bf v})
+ q{\bf A}\Big)\cdot\delta{\bf u}_{s}
- q\delta \Phi +{\bf u}_{s}\cdot\delta {\bf A} \bigg]
\nonumber \\
& &\qquad\qquad\qquad
+\ \delta F\ \left[\frac{1}{2}m|{\bf u}_{s}|^{2}
+\frac{1}{2}m|{\bf u}_{s}-{\bf v}|^{2}-q\Phi
+q{\bf u}_{s}\cdot{\bf A}
\right]\bigg\}
\label{eq:bigvary} \\
& & + \int dt\int d{\bf x}\Big(\nabla _{\bf x} \Phi
+ \, \frac{\partial {\bf A}}{\partial t}\Big) \!\cdot\!
\Big(\nabla _{\bf x}\delta \Phi
+\delta\frac{\partial {\bf A}}{\partial t}\Big)
 - (\nabla_{\bf x}\times{\bf A})
\!\cdot\!(\nabla_{\bf x}\times\delta {\bf A})\,.
\nonumber
\end{eqnarray}
Stationary variations in $\Phi $ and ${\bf A}$ yield:
\begin{eqnarray}
\nabla _{\bf x}\cdot\left(\nabla_{\bf x} \Phi
+\frac{\partial {\bf A}}{\partial t}\right)
= -\, q\int d{\bf v}F({\bf x},{\bf v},t),
\nonumber \\
\nabla_{\bf x}\times(\nabla_{\bf x}\times{\bf A})
=-\,\frac{\partial }{\partial t}\left(\nabla_{\bf x} \Phi
+ \frac{\partial {\bf A}}{\partial t}\right)
+q\int d{\bf v}\,F({\bf x},{\bf v},t) {\bf u}_{s}\,.
\label{Maxwell3}
\end{eqnarray}
Thus, Maxwell's equations for the electromagnetic field of the
plasma are recovered by requiring $\delta l=0$ for all variations
of the field potentials $\Phi$ and {\bf A}. To continue toward the
Euler-Poincar\'{e} form of the Maxwell-Vlasov equations, one must
determine the forms of the variations $\delta {\bf u}_{s}$ and
$\delta F$ in (\ref{eq:bigvary}).

According to the general theory, variations in the particle
evolution map $\psi$ lead to variations in the phase space
velocity $\delta {\bf u}$ of the form
\begin{equation}
\delta {\bf u}=
\frac{\partial {\bf w}}{\partial t}
+ [{\bf u},{\bf w}]
\equiv
\frac{\partial {\bf w}}{\partial t}
- {\rm ad}_{\bf u}
{\bf w}\,.
\label{eq:variation2}
\end{equation}
This Euler-Poincar\'{e} form of the variations may also
be verified by a direct tensorial calculation, which is given in Holm,
Marsden and Ratiu [1997]. The spatial part of this equation gives the
variation of the spatial part of the field ${\bf u}$.

Variations of the field $\psi$ also induce variations of
the density $F$, in the same way as the parameter variations are
induced in the general theory for the Euler-Poincar\'e equations, see
equation (\ref{tensor.eqn}). Either from that equation, or by direct
calculations, these variations are computed to be
\begin{equation}
\delta F = -\nabla_{\bf z}
\cdot (F{\bf w}), \label{varyeff}
\end{equation}
which is equivalent to the formula
$$\delta (Fd{\bf x}d{\bf v})
= -{\pounds}_{\bf w} (Fd{\bf x}d{\bf v}).$$
\paragraph{Computation of the variations.} With these formulae for
$\delta {\bf u}$ and
$\delta F$ in place, we compute
\begin{eqnarray}
\delta \mathfrak{S}_{\rm red} & = & \int dt\int d{\bf x}d{\bf v}F
\left[\Big(m{\bf u}_{s} + m({\bf u}_{s}-{\bf v})
+ q{\bf A}\Big)\!\cdot\!\Big(\frac{\partial }{\partial t}{\bf w}
+ [{\bf u},{\bf w}]
\Big)\right]
\nonumber \\
& &  -\ \nabla_{\bf z}
\cdot (F{\bf w})
\left(\frac{1}{2}(m|{\bf u}_{s}|^{2}
+ m|{\bf u}_{s}-{\bf v}|^{2})
+ q{\bf u}_{s}\cdot {\bf A}-q\Phi \right).
\label{eq:bigvary2}
\end{eqnarray}
Integrating by parts and dropping boundary terms gives
\begin{eqnarray}
    \delta \mathfrak{S}_{\rm red} & = & \int dt\int d{\bf x}d{\bf v}\
       {\bf w}\cdot\bigg[
       -\,\frac{\partial}{\partial t}\left(Fm({\bf u}_{s}
       +({\bf u}_{s}-{\bf v})
       + \frac{q}{m}{\bf A})\right)
\nonumber \\
    & & - \, {\rm ad}^{\ast}_{\bf u}\left(Fm({\bf u}_{s}
        + ({\bf u}_{s}-{\bf v})+\frac{q}{m}{\bf A})\right)
\nonumber \\
    & & + \, F\nabla_{\bf z} \left(\frac{1}{2}m|{\bf u}_{s}|^{2}
        +\frac{1}{2}m|{\bf u}_{s}-{\bf v}|^{2}
         +q{\bf u}_{s}\cdot {\bf A}-q\Phi \right)\bigg].
\label{eq:bigvary3}
\end{eqnarray}
Expanding the ${\rm ad}^{\ast}$ term using formula (\ref{eq:ad1})
results in
\begin{eqnarray}
      \delta \mathfrak{S}_{\rm red} & = &  \int dt \int d{\bf x}
d{\bf v} \ {\bf w}\cdot
        \bigg[ -\frac{\partial F}{\partial t}m \left( {\bf u}_{s}
       + ({\bf u}_{s}-{\bf v})+\frac{q}{m}{\bf A} \right)
\nonumber \\
     & & - \, Fm\frac{\partial }{\partial t}
         \bigg({\bf u}_{s}+({\bf u}_{s}
         -{\bf v}) + \frac{q}{m}{\bf A}\bigg)
\nonumber \\
     & &  -\left({\bf u} \cdot\nabla_{\bf z}\right)
         \bigg(Fm \left( {\bf u}_{s}+
         ({\bf u}_{s}-{\bf v})+\frac{q}{m}{\bf A}\right) \bigg)
\nonumber \\
      & &  - \, Fm\left( {\bf u}_{s}+({\bf u}_{s}-{\bf v})
        +\frac{q}{m}{\bf A} \right) (\nabla_{\bf z} \cdot{\bf u})
           \nonumber \\
       & &  -\left(Fm\left( {\bf u}_{sj}+({\rm u}_{sj}-{\rm v}_{j})
            +\frac{q}{m}{\rm A}_{j}\right) \right)
           \nabla_{\bf z}u^{j}
\nonumber \\
     & &   +  F\nabla_{\bf z} \left(\frac{1}{2}m|{\bf u}_{s}|^{2}
           +\frac{1}{2}m|{\bf u}_{s}-{\bf v}|^{2}
            +q{\bf u}_{s}\cdot{\bf A}-q\Phi \right)\bigg].
  \label{eq:bigvary4}
\end{eqnarray}
We expand the products to obtain
\begin{eqnarray}
\delta \mathfrak{S}_{\rm red} & = &  \int dt\int d{\bf x}d{\bf
v}\;{\bf w}\cdot
           \bigg\{-m\left( {\bf u}_{s}+({\bf u}_{s}-{\bf v})
           +\frac{q}{m}{\bf A}\right) \left(\frac{\partial F}
           {\partial t}
           +{\bf u} \cdot\nabla_{\bf z}F\right)
\nonumber \\
        &  &  -\,Fm\left( {\bf u}_{s}+({\bf u}_{s}-{\bf v})+
        \frac{q}{m}{\bf
           A}\right)  (\nabla_{\bf z} \cdot{\bf u})
\nonumber \\
       &  &  -\,Fm\bigg[ \Big(\frac{\partial}{\partial t}
                        +\big({\bf u}\cdot\nabla_{\bf z}\big)\Big)
          \left({\bf u}_{s}+({\bf u}_{s}-{\bf v})+
          \frac{q}{m}{\bf A}\right)
            +\frac{q}{m}\nabla_{\bf z}\Phi \bigg]
\nonumber \\
        &  & -\,Fm\left( {\rm u}_{sj}+({\rm u}_{sj}-{\rm v}_{j})
             +\frac{q}{m}{\rm A}^{j}\right)
               \nabla_{\bf z}u^{j}
             + Fm{\rm u}_{sj}\nabla_{\bf z}{\rm u}_{s}^{j}
             + Fq{\rm A}^{j}\nabla_{\bf z}{\rm u}_{sj}
\nonumber \\
         &  & +\,Fm({\rm u}_{sj}
              -{\rm v}_{j})\nabla_{\bf z}({\rm u}_{s}^{j}
              -{\rm v}_{s}^{j})
               +Fq{\rm u}_{sj}\nabla_{\bf z}{\rm A}^{j}\bigg\} .
\label{eq:bigvary5}
\end{eqnarray}
Consider the last two lines of equation (\ref{eq:bigvary5}).
Upon writing ${\bf w}
=({\bf w}_{1},{\bf w}_{2})$, where
${\bf w}_{1}, {\bf w}_{2}\in {\mathbb R }^{3}$,
these lines reduce to:
\begin{eqnarray}
& & \quad  -\; Fm({\bf u}_{s}+({\bf u}_{s}-{\bf v}))\cdot
\left( {\bf w}_{1}\cdot \nabla _{\bf x}
+{\bf w}_{2}\cdot \nabla _{\bf v} \right)
{\bf u} \nonumber \\
& & \quad + \; Fm({\bf u}_{s}+({\bf u}_{s} -{\bf v}))\cdot
\left( {\bf w}_{1}\cdot \nabla _{\bf x}
+{\bf w}_{2}\cdot \nabla _{\bf v}\right)
{\bf u}_{s} \nonumber \\
& & \quad -\;Fq{\bf A}\cdot\left({\bf w}_{1}\cdot\nabla_{\bf x}
+{\bf w}_{2}\cdot \nabla_{\bf v}\right)
{\bf u}
+Fq{\bf A}\cdot\left({\bf w}_{1}\cdot\nabla_{\bf x}
+{\bf w}_{2}\cdot
\nabla_{\bf v}\right){\bf u}_{s} \nonumber \\
& & \quad +\;Fq{\bf u}_{s}\cdot\left({\bf w}_{1}\cdot\nabla_{\bf x}
+{\bf w}_{2}\cdot \nabla_{\bf v}\right){\bf A}
-Fm({\bf u}_{s}-{\bf v})\cdot
\left( {\bf w}_{1}\cdot \nabla _{\bf x}
+ {\bf w}_{2}\cdot \nabla _{\bf v}\right) {\bf v} \nonumber \\
& & =  Fq{\bf u}_{s}\left({\bf w}_{1}
\cdot \nabla_{\bf x}\right){\bf A}
- Fm({\bf u}_{s}-{\bf v})\cdot {\bf w}_{2}.
\label{eq:eureka}
\end{eqnarray}
The first three lines cancel to zero because they only involve
spatial velocity projections, where ${\bf u}={\bf u}_{s}$. The last
line follows upon using $\nabla_{\bf x}{\bf v}=0$ and $\nabla_{\bf
v}{\bf A}=0$; which hold, respectively, because ${\bf v}$ is an
independent coordinate and ${\bf A}$ is a function of space alone.
Similarly, and under the additional observation that $\nabla_{\bf
z}\Phi=(\nabla_{\bf x}\Phi,0)$ because the potential $\Phi$ also
does not depend on velocity, the other three lines of equation
(\ref{eq:bigvary5}) are purely spatial, i.e., the projection onto
the last three coordinates would give zero, and hence the
contribution to the variation of the action
$\delta \mathfrak{S}_{\rm red}$ from ${\bf w}_{2}$ comes only from the
calculation in equation (\ref{eq:eureka}). Stationarity of the action
under the velocity components of the variation, ${\bf w}_2$, then
implies:
\begin{equation}
Fm({\bf u}_{s}-{\bf v})=0,
\; \mbox{ i.e., } \;  {\bf u}_{s} ={\bf v}.
\label{well}
\end{equation}
Consequently, in equation (\ref{eq:bigvary5})
we can write ${\bf u}$ as
$({\bf v},{\bf a})$ where {\bf a} is yet to be determined, and we
can also replace ${\bf u}_{s}-{\bf v}$ with zero. On doing this, the
contribution to the variation of the action from ${\bf w}_{1}$
becomes
\begin{eqnarray}
 \delta \mathfrak{S}_{\rm red} & = &    \int dt\int d{\bf x}d{\bf
v}\;
             {\bf w}_{1}\cdot \bigg[-(m{\bf v}+q{\bf A})
            \left( \frac{\partial F}{\partial t}
            +\nabla_{\bf z} \cdot (F{\bf u})\right)
                \nonumber \\
      & & - \, F\bigg(m\frac{\partial {\bf v}}{\partial t}
            +m({\bf v}\cdot \nabla_{\bf x}) {\bf v}
            +m({\bf a}\cdot \nabla_{\bf v}){\bf v} \nonumber \\
       & & + \, q\frac{\partial {\bf A}}{\partial t}
             +q\nabla_{\bf x}\Phi
            + q {\bf v}\times(\nabla_{\bf x}
            \times {\bf A})\bigg) \bigg]\,.
\label{eq:there}
\end{eqnarray}
Here, we have used standard vector
identities in obtaining the result
\begin{equation}
{\bf w}\cdot qF
\left({\rm u}_{sj}\nabla_{\bf z}{\rm A}_{s}^{j}
-({\bf u}\cdot \nabla_{\bf z}){\bf A}\right)
= qF {\bf w}_{1}\cdot\left({\bf v}
\times(\nabla_{\bf x} \times {\bf A})\right).
\label{gosh}
\end{equation}
Referring to the continuity equation (\ref{useful}) for $F$ and using
the identities $\partial{\bf v}/ \partial t = 0$ and
$\nabla_{\bf x}{\bf v} = 0$ reduces equation (\ref{eq:there}) to:
\[
\delta \mathfrak{S}_{\rm red} =
-\int dt\int d{\bf x}d{\bf v} \,{\bf w}_{1}\cdot F\left( m{\bf a}
+ q\nabla_{\bf x}\Phi+q\frac{\partial {\bf A}}{\partial t}
- q{\bf v}\times(\nabla_{\bf x}\times {\bf A}) \right) .
\]
Therefore, $\delta \mathfrak{S}_{\rm red} = 0$ implies that
\begin{equation}
 m{\bf a}=
-\,q\nabla_{\bf x}\Phi-q\frac{\partial {\bf A}}{\partial t}
+ q{\bf v}\times(\nabla_{\bf x}\times{\bf A}).
\label{you}
\end{equation}

Now consider what the invariance of the Boltzmann function $f$
implies. By equation (\ref{eff}) and substitution for
${\bf u}=({\bf v},{\bf a})$, we obtain
\begin{equation}
\frac{\partial f}{\partial t}+{\bf v}\cdot \nabla _{\bf x}f-
\frac{q}{m}\bigg[\left(\nabla _{\bf x}\Phi
+\frac{\partial {\bf A}}{\partial t}\right)-
{\bf v}\times(\nabla_{\bf x}\times {\bf A})\bigg]
\cdot \nabla _{\bf v}f=0,
\label{haveit}
\end{equation}
and so, along with equations (\ref{Maxwell3}), we have recovered
the full Maxwell-Vlasov system from stationarity of the
action (\ref{eq:newlag}) entirely in the Eulerian description.

\section{The Generalized Legendre Transformation.}\label{absleg.sec}
\paragraph{Introduction.} Before passing to the Hamiltonian
description of the Maxwell-Vlasov equations, we pause to explain
the theoretical background of how one does this when there are
degeneracies. This section can be skipped if one is willing to
simply take on faith that one should {\it do the Legendre
transformation slowly and carefully} when there are degeneracies.

As explained in Marsden and Ratiu [1994], one normally
thinks of passing from Euler--Poincar\'e equations on a Lie algebra
$\mathfrak{g}$ to Lie--Poisson equations on the dual
$\mathfrak{g}^\ast$ by means of the Legendre transformation. In
some situations involving the Euler-Poincar\'e equations, one starts
with a Lagrangian on $\mathfrak{g} \times V ^\ast $ and performs a
{\it partial} Legendre transformation, in the variable $\xi$
only, by writing
\begin{equation}\label{legendre}
\mu = \frac{\delta l}{\delta \xi}\,, \quad
h(\mu, a) = \langle \mu, \xi\rangle - l(\xi, a).
\end{equation}
Since
\begin{equation}
\frac{\delta h}{\delta \mu} = \xi +\left \langle \mu, \frac{\delta
\xi}{\delta \mu} \right \rangle - \left \langle \frac{\delta
l}{\delta \xi}\,, \, \frac{\delta \xi}{\delta \mu} \right \rangle\,
= \,\xi\,,
\end{equation}
and $\delta h / \delta a = -\delta l / \delta a$,
we see that the Euler-Poincar\'e equations
(\ref{eulerpoincareright1}) for $\xi\in\mathfrak{g}$ and $\dot a(t) =
-a(t)\xi(t)$ imply the Hamiltonian semidirect-product Lie--Poisson
equations for $\mu\in\mathfrak{g}^\ast$. Namely,
\begin{equation} \label{LPright1}
 \frac{\partial }{\partial t}\mu = -
\mbox {\rm ad}_{({\delta h}/{\delta \mu})}^{\ast} \mu
- \frac{\delta h }{ \delta a} \diamond  a
= \{ \mu, h\}_{LP}\,,
\quad
 \frac{\partial }{\partial t} a = - a \frac{\delta h}{\delta \mu}
= \{ a, h\}_{LP}\,,
\end{equation}
with ($+$) Lie-Poisson bracket on $\mathfrak{g}^\ast\times V^\ast$
given by
\begin{equation} \label{LP-brkt}
\{g, h\}_{LP}
= - \left \langle \mu, {\rm ad}_{({\delta h}/{\delta \mu})}
\frac{\delta g}{\delta \mu}\right \rangle
+\left \langle a,
  \frac{\delta g}{\delta a}\frac{\delta h}{\delta \mu}
- \frac{\delta h}{\delta a}\frac{\delta g}{\delta \mu}
\right \rangle\,.
\end{equation}
If the Legendre transformation (\ref{legendre}) is invertible, then
one can also pass from the Lie--Poisson equations to the
Euler--Poincar\'e equations together with the equations $\dot a(t) =
-a(t)\xi(t)$.

It is important in this paper to give a detailed explanation that
incorporates the degeneracy of the parameter dependent system
together with the role of symmetry. Unlike the examples considered in
Holm, Marsden and Ratiu [1997] such as compressible flow or MHD, in
the case of the Maxwell-Vlasov system or even the Vlasov-Poisson
system, the Lagrangian $L_{a_0}$ corresponding to the action in
equation (\ref{eq:newlag}) is {\it degenerate}, since it does not
depend on the variables $\dot{\Phi}$ and $\dot{{\bf v}}$. In other
words, {\it the degeneracy and corresponding constraints that appear
in Vlasov plasmas are more serious than for fluids or the heavy top,
etc.} To deal with this degeneracy, we shall use the generalized
Legendre transformation in the context of Lagrangian submanifolds, as
described in Tulczyjew [1977]. This is also related to the Dirac
theory of constraints (see Dirac [1950]). In particular, we shall take
special care to ensure that the Hamiltonian formulation of the
Maxwell-Vlasov system preserves the constraints associated with the
degeneracy of its Lagrangian.

\paragraph{The general construction.} Let $Q$ be a manifold and $\pi :
T^{\ast}Q\rightarrow Q$ be the cotangent bundle of $Q$. Then
$TT^{\ast}Q$ is a symplectic manifold with a symplectic form that
can be written in two distinct ways as the exterior derivative of
two intrinsic one forms. These two one forms are denoted $\lambda$
and $\chi$ and are given in coordinates by:
\begin{equation}
\lambda = \dot{p}dq + pd\dot{q}
\end{equation}
and
\begin{equation}
\chi = \dot{p}dq - \dot{q}dp,
\end{equation}
where $(q, p)$ are coordinates for $T^\ast Q$ and
$(q, p, \dot{q}, \dot{p})$ are the corresponding
coordinates for $T T^\ast Q$.
For the intrinsic definitions of these forms, see Tulczyjew [1977].

Let $L : J \rightarrow \mathbb{R}$ be a Lagrangian defined on a
submanifold $J \subset TQ$ called the {\bfi Lagrangian constraint}.
The Legendre transformation is a procedure to obtain a Hamiltonian
$H : K \rightarrow \mathbb{R}$
defined on a submanifold $K \subset T^{\ast}Q$, called the
{\bfi Hamiltonian constraint}. The Euler-Lagrange equations are:
\begin{equation}
\lambda = dL \qquad \mbox{on} \quad J\,,
\label{1eq}
\end{equation}
while the Hamilton equations are
\begin{equation}
\chi = -dH \qquad \mbox{on} \quad K\,.
\label{2eq}
\end{equation}
The abbreviated expressions (\ref{1eq}) and (\ref{2eq}) stand for
\begin{equation}
\lambda = d(L\circ T\pi) \quad \mbox{on } \quad (T\pi)^{-1}(J)\,,
\end{equation}
and
\begin{equation}
\chi = -d(H \circ \tau ^{-1}) \quad \mbox{on } \quad (\tau)^{-1}(K)\,,
\end{equation}
where $\tau$ is the canonical projection $\tau: T T^\ast Q
\rightarrow T^\ast Q$, given in coordinates by
$\tau(q,p,\dot{q},\dot{p}) = ( q,p )$. The map $T \pi $ is given by
$T\pi (q,p,\dot{q},\dot{p}) = (q , \dot{q})$.

Both the Euler-Lagrange
and Hamilton equations define the same Lagrangian submanifold $D$ of
$TT^{\ast}Q$. The Lagrangian and Hamiltonian
$L$ and $H$ are the generating functions with respect to the one
forms $\lambda$ and $\chi$ respectively.

The {\bfi generalized Legendre transformation}
consists of the following steps:
\paragraph{Step 1.} For each
$(q,p) \in T^{\ast}Q$ define
\begin{equation}
K(q,p) =\left\{ (q,\dot{q}) \in T_{q}Q \; \left| \;
\frac{\partial}{\partial \dot{q}}\left(p\dot{q} -
L(q,\dot{q})\right) = 0 \right\} \right. \,,
\end{equation}
and let
\begin{equation}
K = \{(q,p) \in T^{\ast}Q \mid K(q,p) \ne  \varnothing \}\,.
\end{equation}
\paragraph{Assumption.} Assume that for each $(q,p) \in K$, the
submanifold $K(q,p)$ is connected. This implies that the stationary
value
\begin{equation}
{\rm stat}_{\dot{q}}(p\dot{q} - L(q,\dot{q}))
\end{equation}
of $p\dot{q} - L(q,\dot{q})$ on $K(q,p)$ is uniquely defined; that
is, it does not depend on $\dot{q}$.
\paragraph{Step 2.} Define $H : K \rightarrow \mathbb{R}$ as follows:
\begin{equation}
H(q,p) = {\rm stat}_{\dot{q}}(p\dot{q} - L(q,\dot{q})\,.
\end{equation}
\paragraph{The generalized Legendre transformation with parameters
and symmetry.} Now we adapt this methodology to the case of parameter
dependent Lagrangians with symmetry. Let
$L_{a_0} : TG\times T{\mathcal Q}\rightarrow \mathbb{R}$
be a Lagrangian depending on  a parameter
${a_0} \in V^{\ast}$. Assume that $G$ acts on  $V^{\ast}$ on the
right and denote by $ag$ the action of $g \in G$ on $a \in V^{\ast}$.
Assume also the following invariance property:
\begin{equation}  L_{ah}(gh, \dot{g}h, q, \dot{q})
= L_{a}(g, \dot{g}, q, \dot{q} ),
\end{equation}
for all $g, h \in G$, all $(q,\dot{q}) \in T{\mathcal Q}$ and all
$a \in V^{\ast}$. A typical element of
$T^{\ast}G\times T^{\ast}{\mathcal Q}$
will be denoted $(g,\alpha_{g}, q, \nu_{q})$ or simply
$(g,\alpha, q, \nu)$.
For each $a_0 \in V^{\ast}$ and  $(g, \alpha) \in T^{\ast}G$, define
\begin{eqnarray}
K_{a_0}(g, \alpha ,q, \nu) & = & \left\{ (g, \dot{g}, q,\dot{q})
\left|
 \frac{\partial}{\partial \dot{g}}
\Big( \alpha \dot{g} + \nu \dot{q}
- L_{a_0}\left(g, \dot{g},q,\dot{q}\right) \Big) = 0
\right. \right.
\nonumber \\
& & \quad
\mbox{and} \quad \left. \frac{\partial}{\partial \dot{q}}
\Big( \alpha \dot{g} + \nu \dot{q} -
L_{a_0}\left(g, \dot{g},q,\dot{q}\right) \Big) = 0 \right\}\,.
\end{eqnarray}
One can immediately check for any $a_0 \in V^{\ast}$, $h \in G$ and
$(g, \alpha, q, \nu) \in T^{\ast}G\times T^{\ast}{\mathcal Q}$ that
$K_{a_0h}(gh, \alpha h, q, \nu) = K_{a_0}(g, \alpha, q, \nu )h$.
Define
\begin{equation}
K_{a_0} = \{(g, \alpha, q, \nu) \mid K_{a_0}(g, \alpha , q, \nu)
\ne \varnothing \}\,.
\end{equation}
Then one can easily prove for any $h \in G$ that $K_{a_0 h} =
K_{a_0}h$. Define
\begin{equation}
K = \{(g, \alpha, q, \nu, a) \mid
K_{a}(g, \alpha, q, \nu) \ne \varnothing\}\,.
\end{equation}
Then $K \subset T^{\ast}G \times T^{\ast}{\mathcal Q} \times V^{\ast}$
is an invariant subset under the action of $G$ given by
$(g, \alpha, q , \nu, a)h = (gh, \alpha h, q , \nu, ah)$.
Now for each $a_0 \in V^{\ast}$ we define
$H_{a_0} :  K_{a_0} \rightarrow \mathbb{R}$ by
\begin{equation}
H_{a_0}(g, \alpha , q , \nu)
= \alpha \dot{g} + \nu \dot{q}
- L_{a_0}(g, \dot{g} , q , \dot{q})\,,
\end{equation}
for any $(g, \dot{g}, q, \dot{q}) \in K_{a_0}(g, \alpha , q , \nu)$.
Then, according to the general theory explained above, Hamilton's
equations are, for each $a_0 \in V^{\ast}$, $-dH_{a_0} = \chi$ on
$K_{a_0}$, where
\begin{equation}
\chi =
\dot{\alpha} dg - \dot{g} d \alpha + \dot{\nu} dq - \dot {q} d
\nu.\end{equation}
One can also easily prove, using the previous equalities, that
$H_{a_0}(g, \alpha , q , \nu)$ has the following invariance property,
\begin{equation}
H_{a_0h}(gh, \alpha h, q , \nu) =  H_{a_0}(g, \alpha , q , \nu).
\end{equation}
Let $\mathfrak{s}^{\ast}$ be the dual of the semidirect
product Lie  algebra $\mathfrak{s} = \mathfrak{g} \circledS V$.
Then define
$\mathcal{K}\subset\mathfrak{s}^{\ast}\times T^{\ast}{\mathcal
Q}$ by
$$ \mathcal{K}   = \{(\alpha, q , \nu , a) \in \mathfrak{s}^{\ast}
\times T^{\ast}{\mathcal Q} \mid (e, \alpha, q , \nu, a) \in K\}\,,$$
and the Hamiltonian
$h_{\mathcal{K}} : \mathcal{K}  \rightarrow \mathbb{R}$ by
$h_{\mathcal{K}}(\alpha, a , q , \nu) = H_{a}(e, \alpha , q , \nu)$.
Thus, $h_{\mathcal{K}}$ is the restriction to $\mathcal{K}  \subset
\mathfrak{s}^{\ast}$ of the right invariant Hamiltonian $H : K
\rightarrow \mathbb{R}$ given by $H(g, \alpha, q , \nu , a) =
H_{a}(g, \alpha , q , \nu)$. Then, by a natural generalization of
semidirect product theory to include constrained Hamiltonian systems,
we have that Hamilton's equations on $\mathcal{K}  \subset
\mathfrak{s}^{\ast}$ generated by $h_{\mathcal{K}}$ give the
evolution of the system on $\mathcal{K}$ determined by the
Poisson-Hamilton equations
$\dot{f} = \{f, h_{\mathcal{K}}\}$ on the Poisson submanifold
$\mathcal{K}  \subset \mathfrak{s}^{\ast} \times T^{\ast}{\mathcal
Q}$, where the Poisson structure is defined in a natural way.
More precisely, we have the Dirac brackets on $K$
(see for instance Dirac [1950] or Marsden and Ratiu [1994])
which, by reduction, give the brackets on $\mathcal{K}$. This is
the abstract procedure underlying the computations we do in the
specific case of plasmas given in the next section.

\section{Hamiltonian formulation} \label{sec-Ham}

We now pass to the corresponding Hamiltonian formulation of
the Maxwell-Vlasov system (\ref{eq:CBE})-(\ref{Maxwell}) in the
Eulerian description by taking the Legendre transform of the
reduced action (\ref{eq:newlag}).

\paragraph{The role of the general theory.} From the geometrical
point of view, we simply apply the generalized Legendre
transformation described abstractly in
\S\ref{absleg.sec} to the degenerate Lagrangian
\[
L_{f_0,\varphi_0}(\psi,\dot{\psi}, \Phi, \dot{\Phi},
{\bf A},\dot{{\bf A}})
:T{\cal F}\times T{\cal V}\times T{\cal A}
\rightarrow\mathbb{R}.
\]
This Lagrangian is degenerate because it does not
depend on the variables $\dot{\Phi}$ and $\dot{{\bf v}}$. The theory
described in \S\ref{absleg.sec} may be applied to this action on
$T({\cal F}\times {\cal V}\times {\cal A})$. The action of the group
${\rm Diff}(T\mathbb{R}^3)$ on the factor ${\cal F}$ for this
Lagrangian is given as before, while the actions on the factors
${\cal V}$ and ${\cal A}$ are trivial. It is easy to see that
the Hamiltonian constraint for each $f_0$ is $K_{f_0} \subset
T^{\ast}({\rm Diff}(T\mathbb{R}^3)\times {\cal V} \times {\cal A})$,
defined by the conditions
\[
  \Psi = \frac{\delta L }{\delta \dot{\Phi}} = 0 \quad  \mbox{and}
  \quad {\bf m}_{v} = \frac{\delta L }{\delta \dot{\bf v}} = 0.
\]
These conditions impose constraints, which for consistency must be
dynamically preserved.
\paragraph{Calculation of the transformed equations.} We will
perform the calculations in detail, working with the reduced
Lagrangian rather than the Lagrangian
$L_{f_0,\varphi_0}(\psi,\dot{\psi}, \Phi, \dot{\Phi},
 {\bf A}, \dot{{\bf A}})$
and setting $\varphi_0 = 1$ as usual.

We start with the action
(\ref{eq:newlag}) for the Maxwell-Vlasov system in the Eulerian
description,
\begin{align}
\mathfrak{S}_{\rm red} ({\bf u}, \Phi , \dot{ \Phi } , {\bf A} ,
\dot{ {\bf A} })
& =  \nonumber \\
& \int\!\! dt\!\!\int\!\! d{\bf x}d{\bf v}
    \; F({\bf x},{\bf v},t) \left(\frac{1}{2}m|{\bf u}_{s}|^{2}
    +\frac{1}{2}m|{\bf u}_{s}-{\bf v}|^{2}
     -q\Phi +q{\bf u}_{s}\cdot{\bf A}\right)
    \nonumber \\
& \quad +\ \frac{1}{2}\int dt\int d{\bf x}\ \Big|\nabla _{\bf x}
   \Phi + \frac{\partial {\bf A}}{\partial t} \Big|^{2}
   - |\nabla_{\bf x}\times {\bf A}|^{2}.
\label{eq:newlag2}
\end{align}
This leads immediately to
\begin{equation}
\frac{\delta l}{\delta \dot{\bf A}}
=\nabla_{\bf x}\Phi + \frac{\partial {\bf A}}{\partial t}= -\, {\bf E},
\label{EandA}
\end{equation}
and so (minus) the electric field variable {\bf E} is the field
momentum density canonically conjugate to the magnetic potential.
Let us define the material momentum density in six dimensions,
\begin{equation}
{\bf m} \equiv \frac{ \delta l }{ \delta {\bf u} }.
\end{equation}
We write  ${\bf m}=({\bf m}_{s},{\bf m}_{v})$, where ${\bf m}_{s}$
is the projection of ${\bf m}$ onto the first three cordinate
positions, and ${\bf m}_{v}$ is the projection onto the last three
places. We think of ${\bf m}_{s}$ and ${\bf m}_{v}$ also as vectors
in six dimensions. From the Lagrangian we see that
\begin{equation}
{\bf m}_{s}=F\Big(m{\bf u}_{s}+m({\bf u}_{s}
-{\bf v})+q{\bf A}\Big)
\quad\hbox{and}\quad
{\bf m}_{v}=0.
\label{moment-def}
\end{equation}

Proceeding with the Legendre transform
of our action (\ref{eq:newlag2}) results in a
corresponding (reduced) Hamiltonian
function written in terms of the velocities,
\begin{eqnarray}
h & = & \int d{\bf x}d{\bf v}\;
F\Big( m|{\bf u}_{s}|^{2}-\frac{1}{2}m|{\bf v}|^{2}+q\Phi\Big)
+{\bf m}_{v}\cdot{\bf a}
\nonumber \\
& & + \;\frac{1}{2}\int d{\bf x}
\Big( |{\bf E}|^{2}
+ |\nabla_{\bf x}\times{\bf A}|^{2}
+ 2{\bf E}\cdot\nabla_{\bf x}\Phi \Big) \,,
\label{Hamil}
\end{eqnarray}
where {\bf a} denotes the projection of
${\bf u}$ onto its last three entries.
Transforming this to the momentum variables gives
\begin{eqnarray}
h & = &  \int d{\bf x}d{\bf
v}\;\frac{1}{4Fm}|{\bf m}_{s}+mF{\bf v}-qF{\bf A}|^{2}
-\frac{1}{2}mF|{\bf v}|^{2}+qF\Phi+{\bf m}_{v}
\cdot {\bf a}
\nonumber \\
& & + \;\frac{1}{2}\int d{\bf x}
\Big( |{\bf E}|^{2}
+ |\nabla_{\bf x}\times{\bf A}|^{2}
+ 2{\bf E}\cdot\nabla_{\bf x}\Phi \Big) \,.
\label{Hamil2} \end{eqnarray}
The variation of this Hamiltonian with respect to
${\bf m},{\bf a},
{\bf E},{\bf A},F$ and $\Phi$ is given by
\begin{eqnarray}
\delta h & = & \int d{\bf x}d{\bf v}\;\bigg[
{\bf u}
\cdot\delta{\bf m}
+{\bf m}_{v}\cdot\delta {\bf a}
-q F{\bf u}_{s}\cdot\delta{\bf A}
+qF\delta \Phi
\nonumber \\
& & - \; \bigg(\frac{1}{2}m|{\bf u}_{s}|^{2}
+\frac{1}{2}m|{\bf u}_{s}-{\bf v}|^{2}
+q{\bf u}_{s}\cdot{\bf A}
-q\Phi\bigg)\delta F\bigg]
\nonumber \\
& & + \; \int d{\bf x}\,
({\bf E}+\nabla_{\bf x}\Phi)\cdot\delta{\bf E}
-(\nabla_{\bf x}\cdot {\bf E})\delta\Phi
+\nabla_{\bf x}\times(\nabla_{\bf x}
\times {\bf A})\cdot\delta {\bf A}.
\label{hamvary1}
\end{eqnarray}
This expression allows one to read off the evolution equations
for the electromagnetic field:
\begin{eqnarray}
\frac{\partial {\bf A}}{\partial t}
\!\!\!\!& = &\!\!\!\! -\,\frac{\delta h}{\delta {\bf E}}
=-\,{\bf E}-\nabla_{\bf x}\Phi, \ \mbox{ i.e., } \
{\bf E}=-\nabla_{\bf x}\Phi-\frac{\partial {\bf A}}{\partial t},
\nonumber \\
\frac{\delta h}{\delta \Phi}
\!\!\!\!& = &\!\!\!\! 0
=  -\nabla_{\bf x}\cdot {\bf E}+q\int d{\bf v}F,
\ \mbox{ i.e., } \  \nabla_{\bf x}\cdot{\bf E}
= q\int d{\bf v}F:=\rho,
\nonumber \\
\frac{\partial {\bf E}}{\partial t}
\!\!\!\!& = &\!\!\!\!
\frac{\delta h}{\delta{\bf A}}
=\nabla_{\bf x}\times(\nabla_{\bf x}\times{\bf A})
-q\int d{\bf v}F{\bf u}_{s},
\ \mbox{ i.e., } \
\frac{\partial {\bf E}}{\partial t}
=\nabla_{\bf x}\times{\bf B}-{\bf j}\,.
\label{maqxwellagain}
\end{eqnarray}
Note that the constraint $ \delta h / \delta \Phi = 0 $ (Gauss' law)
arises from the absence of $\dot{\Phi}$ dependence in $l$.

The general theory of \S\ref{absleg.sec} shows that $F$ is an
element of the second factor of the semidirect product and so its
evolution is given by Lie dragging as a density. Likewise, $f$ is
Lie dragged as a scalar and $ m_i$ satisfies a Lie-Poisson
evolution equation:
\begin{eqnarray}
\frac{\partial F}{\partial t}
& =&-\nabla_{\bf z}
\cdot(F{\bf u}),
\nonumber \\
\frac{\partial f}{\partial t}
&=&-{\bf u}
\cdot \nabla_{\bf z}f,
\nonumber \\
\frac{\partial m_i}{\partial t}
&=&-\,\frac{\partial }{\partial z^{j}}m_iu^{j}
-m_j\frac{\partial }{\partial z^{i}}u^{j}
-F\frac{\partial}{\partial z^{i}}\frac{\delta h}{\delta F}.
\label{LPeqn}
\end{eqnarray}
The first two of these equations reflect the assumptions that were
made in the definitions of $f$ and $F$, whilst the last equation
encodes the dynamics for the system. We first consider the case
where the momentum component $i$ takes the values $4,5,6$. In this
case,
\begin{eqnarray}
-\,\frac{\partial m_i}{\partial t}
\!\!\!& = &\!\!\!
{\rm m}_{sj} \frac{\partial }{\partial z^{i}}u^{j}
+ {\rm m}_{vj}\frac{\partial }{\partial z^{i}}u^{j}
- F\frac{\partial }{\partial z^{i}}
\Big(\frac{1}{2}m|{\bf u}_{s}|^{2}
\!+\! \frac{1}{2}m|{\bf u}_{s}-{\bf v}|^{2}
\!+\! q{\bf u}_{s}\cdot{\bf A}-q\Phi\Big)
\nonumber \\
\!\!\!& = &\!\!\!
{\rm m}_{vj}\frac{\partial }{\partial z^{i}}u^{j}
+ Fm({\rm u}_{sj}
- {\rm v}_{j})\frac{\partial u^{j}}{\partial z^{i}}
- Fm({\rm u}_{sj}
- {\rm v}_{j})\frac{\partial }{\partial z^{i}}({\rm u}_{s}^{j}
- {\rm v}_{s}^{j})
\nonumber \\
& & - \; qF{\rm u}_{sj}\frac{\partial {\rm A}_{s}^{j}}{\partial z^{i}}
+ Fq\frac{\partial }{\partial z^{i}}\Phi,
\label{emvee}
\end{eqnarray}
where $i = 4,5,6$. In the second line of
equation (\ref{emvee}), we have substituted for
${\bf m}_{s}$ from equation (\ref{moment-def}) and rearranged terms.
Here ${\bf m}_v = 0$, because $l$ does not depend on
$\dot{{\bf v}}$. Setting ${\bf m}_{v}=0$ initially in equation
(\ref{emvee}) ensures that ${\bf m}_{v}\equiv 0$ persists throughout
the ensuing motion; for potentials $\Phi$ and ${\bf A}$ that are
independent of {\bf v}, and provided the constraint holds that ${\bf
u}_{s}={\bf v}$, as in equation (\ref{well}). Likewise, the Gauss'
law constraint imposed by $\delta h/\delta \Phi = 0$ also persists
during the ensuing motion, as seen from the last equation of
(\ref{maqxwellagain}) and the first equation of (\ref{LPeqn}),
provided the constraint ${\bf u}_{s}={\bf v}$ holds and $F$ vanishes
in the limit as $|{\bf v}|\to\infty$ .

The spatial part of the evolution equation of
${\bf m}$ will produce the required single-particle
dynamics. From equation (\ref{LPeqn}), we have
\begin{eqnarray}
\frac{\partial m_i}{\partial t}
= -\,\frac{\partial}{\partial z^{j}}m_iu^{j}
- m_j\frac{\partial}{\partial z^{i}}u^{j}
- F\frac{\partial }{\partial z^{i}}\frac{\delta h}{\delta F}.
\label{meqn}
\end{eqnarray}
Setting $i=1,2,3,$ in equation (\ref{meqn}), then substituting for
$\delta h/\delta F$ and using the relations
\[ {\bf m}_{s}=F(m{\bf u}_{s}+m({\bf u}_{s}-{\bf v})
+q{\bf A}), \quad
{\bf m}_{v}\equiv 0, \]  and
\[ \nabla _{\bf v} \Phi =0,\quad  \nabla _{\bf v} {\bf A} =0,\]
yields the spatial components of the motion equation,
\begin{eqnarray}
\frac{\partial {\rm m}_{si}}{\partial t} & = &
-\,\frac{\partial }{\partial z^{j}}{\rm m}_{si}u^{j}
- {\rm m}_{sj}\frac{\partial }{\partial z^{i}}u^{j} \nonumber \\
& & - \,  F\frac{\partial }{\partial z^{i}}
\bigg(\frac{1}{2}m|{\bf u}_{s}|^{2}
+ \frac{1}{2}m|{\bf u}_{s}-{\bf v}|^{2}
+q{\bf u}_{s}\cdot{\bf A}-q\Phi\bigg)\,.
\label{meqn2}
\end{eqnarray}
Substituting for ${\bf m}_{s}$ and then using the continuity relation
${\partial}F/{\partial t}+\nabla_{\bf z}\cdot(F{\bf u})=0$
gives
\begin{eqnarray}
m\frac{\partial {\rm u}_{si}}{\partial t}+
q\frac{\partial {\rm A}_{si}}{\partial t  }
\!\!& = &\!\!
-\ u^{j}\frac{\partial }{\partial z^{j}}m{\rm u}_{si}
-{\rm u}_{s}^{j}\frac{\partial }{\partial z^{j}} q{\rm A}_{si}
-q{\rm A}_{j}\frac{\partial }
{\partial z^{i}}u^{j}-q\frac{\partial \Phi}{\partial z^{i}}\nonumber \\
&&+ \, q\frac{\partial }{\partial z^{i}}({\bf u}_{s} \cdot {\bf A})
-\frac{1}{2}m\frac{\partial }{\partial z^{i}}|{\bf u}_{s}
-{\bf v}|^{2}.
\label{meqn3} \end{eqnarray}
Rearranging this equation results in
\begin{eqnarray}
m\bigg(\frac{\partial }{\partial t}+u^{j}
\frac{\partial }{\partial z^{j}}\bigg){\bf u}_{s}=q{\bf E}
+q{\bf u}_{s}\times (\nabla_{\bf x}\times {\bf A})
-\frac{1}{2}m\frac{\partial }{\partial z^{i}}
|{\bf u}_{s}-{\bf v}|^{2}.
\label{lorentz}
\end{eqnarray}
We may now evaluate this on the constraint set ${\bf u}_{s} ={\bf v}$
and thereby obtain the Lorentz force,
\begin{eqnarray}
m{\bf a}=q({\bf E}+{\bf v}\times{\bf B}),
\label{lorentz2}
\end{eqnarray}
where {\bf a} is the acceleration of a fluid parcel (the last three
components in ${\bf u}={\bf v}{\bf a}$). As we have seen, in this
Hamiltonian formulation of the Maxwell-Vlasov  equations in the
Eulerian description, the acceleration ${\bf a}$ in ${\bf u}$ is a
vector Lagrange multiplier which imposes ${\bf m}_{v}=0$. Equation
(\ref{lorentz2}) provides an expression for this Lagrange multiplier
in terms of known dynamical variables and as a consequence we regain
the equation for the acceleration of a charged particle in an
electromagnetic field. The momentum constraint ${\bf m}_{v}=0$
remains invariant when the electromagnetic potentials are independent
of the phase space velocity coordinate {\bf v} and the
velocity constraint ${\bf u}_{s} ={\bf v}$ holds. Perhaps not
unexpectedly, one finds that
$\nabla_{\bf z}\cdot{\bf u}=0$. Also, (minus) the electric field is
canonically conjugate to the vector potential, and the electrostatic
potential $\Phi$ plays the role of a Lagrange multiplier which imposes
Gauss's law. Thus, our Hamiltonian formulation augments the usual
Maxwell-Vlasov description of plasma dynamics by self-consistently
deriving the particle acceleration by the Lorentz force $m{\bf
a}=q({\bf E}+{\bf v}\times {\bf B})$ instead of assuming it {\it a
priori}.

\paragraph{The Poisson Hamiltonian structure.} The general theory
outlined briefly in \S\ref{absleg.sec} also leads to the Poisson
bracket structure for the Maxwell-Vlasov theory on the Hamiltonian
side. However, our Hamiltonian description has a redundancy, namely
the information for the particle trajectories can be recovered from
the spatial plasma density. Explicitly, if we let
$ H(f) = (1/2)|{\bf v}|^2 + \Phi_f({\bf x})$  be the single
particle Hamiltonian determined by the plasma density $f$, then the
flow of this Hamiltonian function can be identified with the
particle evolution map $\psi$. We can also think of this as a
constraint on the level of equations of motion, as the Hamiltonian
vector field of $ H(f)$  must equal the time derivative of the  map
$\psi$, i.e., the particle velocity field in phase space. In other
words, as is well known, the particle dynamics is completely
determined by the plasma density dynamics. This may be regarded as a
constraint on the system that leads to the elimination of the forward
map as a dynamical variable. This ``redundancy'' is of course one of
the sources of degeneracy of the Lagrangian and Hamiltonian structures.

Thus, the constraint explicitly enforcing this consistency condition
leads to a further ``reduction'' which again may be handled by the
Dirac theory of constraints to arrive at the Hamiltonian structure
in terms of the variables $F$ (or equivalently $f$ in view of
the canonical nature of the particle transformations) and the
electromagnetic potentials. The resulting Poisson bracket structure
is given by the Lie-Poisson structure for the $f$'s plus the canonical
structure for the electromagnetic potentials, which was the starting
point for Marsden and Weinstein [1982], who carried out the reduction
of this bracket with respect to the action of the electromagnetic
gauge group to obtain the final Maxwell-Vlasov bracket on the space
with variables $f$, ${\bf E}$ and ${\bf B}$. This procedure was
motivated by and corrected a bracket found by ad hoc methods in
Morrison [1980]. We need not repeat this construction.

\paragraph{The Kelvin-Noether theorem.} A final result worth
mentioning is a Kelvin's theorem for the Maxwell-Vlasov particle
dynamics. These dynamics, given in the last equation in (\ref{LPeqn}),
may be rewritten as
\begin{equation}
\bigg(\frac{\partial }{\partial t}
+\pounds_{\bf u}\bigg)
\bigg(\frac{1}{F}m_i dz^i\bigg)
+ d\frac{\delta h}{\delta F}
= 0,
\label{kelpart}
\end{equation}
so that
\begin{equation}\label{kelvinpl.eqn}
\frac{d}{dt}\oint_{\gamma (t)}
\frac{1}{F}m_i dz^i=0,
\end{equation}
for a loop $\gamma (t)$ which follows the particle trajectories in
phase space. The Kelvin circulation integral in phase space,
\begin{equation}
I=\oint_{\gamma (t)}\frac{1}{F}
m_i dz^i,
\end{equation}
may be evaluated on the invariant constraint manifold ${\bf m}_{v} =0$
as
\begin{equation}
I=\oint_{\gamma (t)}(mu_{si}+qA_i) dx^i.
\end{equation}
We recognize this integral as the {\it Poincar\'{e} invariant} for the
single particle motion in phase space.

The above result follows from the abstract Kelvin-Noether theorem by
letting $\mathcal C := \{
\gamma : S^1 \rightarrow T \mathbb{R}^3  \mid \gamma
{\rm~continuous} \}$ be the space of continuous loops in single
particle velocity phase space and letting the group
${\rm Diff}(T \mathbb{R}^3)$ act on
$\mathcal C$ on the right by
$(\eta, \gamma)\in \mbox {\rm Diff}(T \mathbb{R}^3) \times
{\mathcal C} \mapsto \gamma \circ \eta \in {\mathcal C}$. The
quantity $ \mathcal{K} $ is chosen to be
\begin{equation}
    \left\langle \mathcal{K} (\gamma, F), a \right\rangle   =
    \oint _\gamma \frac{1}{F} a\,.
\end{equation}
The abstract Kelvin-Noether theorem for the Maxwell-Vlasov
equations in Euler-Poincar\'e form then reproduces the version of
Kelvin's theorem given in (\ref{kelvinpl.eqn}).

\section{Conclusion} \label{sec-conc}

In this paper we have cast Low's mixed Eulerian-Lagrangian
action principle for Maxwell-Vlasov theory into a purely Eulerian
description. In this description we find that Maxwell-Vlasov
dynamics are governed by the Euler-Poincar\'{e} equations for
right invariant motion on the diffeomorphism group of $\mathbb{R}^{n}$
($n=6$ for three dimensional Maxwell-Vlasov motion). These
equations were recently discovered by Holm, Marsden and Ratiu [1997]
who investigated the class of Hamilton's principles which are right
invariant under the subgroup of the diffeomorphisms which leaves
invariant a set ${\cal T}$ of tensor fields in the Eulerian
variables. The Maxwell-Vlasov motions invariant under this
subgroup are the steady Eulerian solutions, which, thus, are
identified as relative equilibria. This identification of
steady Eulerian Maxwell-Vlasov solutions as right invariant
equilibria places these solutions into the Hamiltonian
framework required for investigating their nonlinear stability
characteristics using, e.g., the energy-Casimir method (see Holm
et al. [1985]). It was this stated goal that first motivated
Low to write his Lagrangian for Maxwell-Vlasov dynamics.

Thus, our formulation of a purely Eulerian action principle and
its associated Euler-Poincar\'e equations and Hamiltonian
framework advances Low's original intention of using
his action principle for studying stability of plasma equilibria
by placing the entire Maxwell-Vlasov equations (including the
particle dynamics, field dynamics and probability distribution
dynamics) into one self-consistent Hamiltonian picture in the
Eulerian description. (As we discussed, Low used mixed aspects of
both Eulerian and Lagrangian phase space descriptions in his
action principle.)

Our Eulerian Hamilton's principle for Maxwell-Vlasov
dynamics is constrained, and
all of the corresponding Lagrange multipliers have been resolved.
This Hamilton's principle is thus available for further
approximations, e.g., by Hamilton's principle asymptotics (see,
e.g., Holm [1996]).

In summary, we have taken an existing action, due to Low
[1958], for the Maxwell-Vlasov system of equations and
demonstrated how to rederive this system as Euler-Poincar\'{e}
equations. The Euler-Poincar\'{e} form emerges from Hamilton's
principle for a system whose configuration space is a group and
whose action is right invariant under a subgroup. This situation
commonly appears in the Eulerian description of continuum
mechanics. In the case of continuum mechanics, the dynamics takes
place on the group of diffeomorphisms and the Eulerian variables
are invariant under a subgroup of the diffeomorphism group. (This
subgroup corresponds to steady Eulerian flows with non-zero
velocity and vorticity.) We showed that this situation also occurs for
the Maxwell-Vlasov equations of plasma dynamics in the Eulerian
description, by showing that the variations considered take the
appropriate form, and then deriving the Maxwell-Vlasov equations
from the Hamilton's principle for the right invariant action
(\ref{eq:newlag}) in Eulerian variables. We then passed to the
Hamiltonian formulation of this system and found its Lie-Poisson
structure.

As discussed in the introduction, the Euler-Poincar\'{e} form of the
dynamics is naturally adapted for applying Lagrange-D'Alembert methods
for geometrical constraints and control as in Bloch et al. [1996]. In
future work, our Euler--Poincar\'e form of the Maxwell-Vlasov system
shall be implemented to describe the control features of a plasma
driven by an external antenna, following the lines of inquiry begun
in the oscillation center approximation for plasmas by Similon et al.
[1986].

\section{Acknowledgments}

We would all like to extend our gratitude to T. Ratiu for his time
and invaluable input. Some early (unpublished) work by J. Marsden,
P. Morrison and H. Gumral on aspects of the problem addressed in
this paper helped us in the formulation given here and is gratefully
acknowledged. In addition, M. Hoyle would like to thank Dion Burns,
Ion Georgiou, Don Korycansky, Shinar Kouranbaeva, Myung Kim and
David Schneider for their time and the discussions which helped him
understand this material. Work by D. Holm was conducted under the
auspices of the US Department of Energy, supported (in part) by funds
provided by the University of California for the conduct of
discretionary research by Los Alamos National Laboratory.


\subsection*{References}

\begin{description}

\item Arnold, V.I. [1966a]
Sur la g\'{e}ometrie differentielle
des groupes de Lie de dimenson
infinie et ses applications \`{a}
l'hydrodynamique des fluids parfaits.
{\it Ann. Inst. Fourier, Grenoble\/} {\bf 16}, 319--361.

\item Batt, J. and G. Rein [1993]
A rigorous stability result for the Vlasov-Poisson
system in three dimensions,
{\it Ann. Mat. Pura Appl.\/} {\bf 164}, 133--154.

\item Batt, J., P.J. Morrison and G. Rein [1995]
Linear stability of stationary solutions of the Vlasov-Poisson
system in three dimensions. {\it Arch. Rational Mech. Anal.} {\bf 130},
163--182.

\item Binney, J. and S. Tremaine S. [1987]
{\it Galactic Dynamics}, 1st Edition, Princeton University Press.

\item Bloch, A.M., P.S. Krishnaprasad, J.E. Marsden, and R. Murray
[1996] Nonholonomic mechanical systems with symmetry, {\it Arch.
Rat. Mech. An.}, {\bf 136}, 21--99.

\item Cendra, H. and J.E. Marsden [1987]
Lin constraints, Clebsch potentials and variational principles,
{\it Physica D\/} {\bf 27}, 63--89.

\item Cendra, H., A. Ibort, and J.E. Marsden [1987]
Variational principal fiber bundles: a geometric
theory of Clebsch potentials
and Lin constraints, {\it J. Geom. Phys.\/} {\bf 4}, 183--206.

\item Chandrasekhar, K. [1977]
{\it Ellipsoidal Figures of Equilibrium}, Dover.

\item Dewar, R.L. [1972]
A Lagrangian theory for non-linear wave packets in a collisionless
plasma, {\it J. Plasma Physics\/} {\bf 7}, Part 2, 267-284.

\item Dirac, P.A.M. [1950] Generalized Hamiltonian mechanics,
{\it Canad. J. Math.\/} {\bf 2}, 129--148.

\item Ebin, D.G. and J.E. Marsden [1970]
Groups of diffeomorphisms and the motion of an incompressible
fluid, {\it Ann. Math.\/} {\bf 92}, 102--163.

\item Galloway, J.J. and Kim, H. [1971]
Lagrangian approach to non-linear wave interaction in a warm plasma,
{\it J. Plasma Physics\/} {\bf 6} 53.

\item Holm, D.D., J.E. Marsden, T.S. Ratiu, and A. Weinstein [1985]
Nonlinear stability of fluid and plasma equilibria,
{\it Phys. Rep.\/} {\bf 123}, 1--116.

\item Holm, D.D. [1996] Hamiltonian balance equations,
{\it Physica D}, {\bf 98} 379-414.

\item Holm, D. D., Marsden, J. E. and Ratiu,
T. [1997] Euler-Poincar\'{e} equations and semidirect products with
applications to continuum theories {\it In preparation}.

\item Iw\'{i}nski, Z.R. and L.A. Turski [1976]
Canonical theories of systems interacting electromagnetically,
{\it Letters in Applied and Engineering Sciences\/}
{\bf 4}, 179--191.

\item Jeans, J. H. [1902]
The stability of a spherical nebula.
{\it Phil. Trans. Roy. Soc.\/} {\bf 199}, 1--53.

\item Kaufman, A. N. and Dewar, R. L. [1984]
Canonical derivation of the Vlasov-Coulomb noncanonical Poisson structure,
{\it Cont. Math. AMS\/} {\bf 28}, 51--54.

\item Low, F.E. [1958]
A Lagrangian formulation of the Boltzmann--Vlasov equation for
plasmas,  {\it Proc. Roy. Soc. Lond. A\/} {\bf 248}, 282--287.

\item Marsden, J.E. and T.S. Ratiu [1994]  {\it Introduction to
Mechanics and Symmetry.\/} Texts in Applied Mathematics, {\bf  17},
Springer-Verlag.

\item Marsden, J.E. and J. Scheurle [1993a]
Lagrangian reduction and the double spherical pendulum,
{\it ZAMP\/} {\bf 44}, 17--43.

\item Marsden, J.E. and J. Scheurle [1993b]
The reduced Euler-Lagrange equations,
{\it Fields Institute Comm.\/} {\bf 1}, 139--164.

\item Marsden, J.E. and A. Weinstein [1982]
The Hamiltonian structure of the Maxwell-Vlasov equations,
{\it Physica D\/} {\bf 4}, 394--406.

\item Marsden, J.E., A. Weinstein, T.S. Ratiu,
R. Schmid, and R.G. Spencer [1983]
Hamiltonian systems with symmetry, coadjoint
orbits and plasma physics, in
Proc. IUTAM-IS1MM Symposium on
{\it Modern Developments in Analytical Mechanics\/},
Torino 1982, {\it Atti della Acad. della Sc. di Torino\/}
{\bf 117}, 289--340.

\item Morrison, P.J. [1987]
Variational principle and stability of
nonmonotone Vlasov-Poisson equilibria.
{\it Z. Naturforsch.\/} {\bf 42a}, 1115--1123.

\item Morrison, P.J. [1980]
The Maxwell-Vlasov equations as a continuous Hamiltonian system,
{\it Phys. Lett. A\/} {\bf 80}, 383--386.

\item Morrison, P.J. and Pfirsch, D [1989]
Free-energy expressions for Vlasov equilibria,
{\it Phys. Rev A\/} {\bf 40}, 3898-3910.

\item Poincar\'{e}, H. [1890] {\it Th\'eorie des tourbillons},
Reprinted by \'Editions Jacques Gabay, Paris.

\item Poincar\'{e}, H. [1901a] Sur la stabilit\'{e}
de l'\'{e}quilibre des figures piriformes affect\'{e}es par une
masse fluide en rotation, {\it Philosophical Transactions A\/} {\bf
198}, 333--373.

\item Poincar\'{e}, H. [1901b] Sur une forme nouvelle des
\'{e}quations de la m\'{e}chanique, {\it CR Acad. Sci.\/} {\bf 132},
369--371.

\item Similon, P.L., A.N. Kaufman and D.D. Holm [1986]
Oscillation center theory and pondermotive
stabilization of the low-frequency plasma modes.
{\it Phys. Fluids} {\bf 29} (1986) 1908--1922.

\item Sturrock, P.A. [1958]
A variational principle and an energy theorem for small
amplitude disturbances of electron beams and of electron plasmas,
{\it Ann. of Phys.\/} {\bf 4}, 306--324.

\item Tulczyjew, W.M. [1977]
The Legendre transformation,
{\it Ann. Inst. Poincar\'{e}\/} {\bf 27}, 101--114.

\item Wan, Y.H. [1990] Nonlinear stability of stationary spherically
symmetric models in stellar dynamics. {\it Arch. Rational Mech.
Anal.} {\bf 112}, 83--95 and preprint, [1997].

\item Ye, H. and P.J. Morrison [1992]
Action principles for the Vlasov equation.
{\it Phys. Fluids B\/} {\bf 4}, 771--777.

\end{description}

\end{document}